\documentclass[12pt]{article}

\usepackage{amsmath,amssymb,amsfonts,graphics,graphicx,amscd,amsfonts,epsf,epsfig,color,epstopdf,url}
\def\m{\mu}

\allowdisplaybreaks[4]

\date{}
\begin{document}

\begin{flushright}

\end{flushright}

\vspace{0.1cm}

\begin{center}
  {\Large

Entanglement and Confinement
in Coupled Quantum Systems

  }
\end{center}
\vspace{0.1cm}
\vspace{0.1cm}
\begin{center}

Fabien Alet,$^a$ Masanori Hanada,$^b$ Antal Jevicki$^c$ and Cheng Peng$^{c,d,e}$\\
\vspace{0.5cm}

$^a${\it Laboratoire de Physique Théorique, Université de Toulouse, CNRS, UPS, France}\\
  \vspace{0.2cm}
$^b${\it STAG Research Centre, University of Southampton, Southampton, SO17 1BJ, UK}\\
\vspace{0.2cm}
$^c${\it Kavli Institute for Theoretical Sciences (KITS) and CAS Center for Excellence in Topological Quantum Computation, University of Chinese Academy of Sciences, Beijing 100190, China }\\
$^d${\it Department of Physics, Brown University, 182 Hope Street, Providence, RI 02912, USA}\\
\vspace{0.2cm}
$^e${\it Center for Quantum Mathematics and Physics (QMAP), Department of Physics\\
University of California, Davis, CA 95616 USA}\\

\end{center}

\vspace{1.5cm}

\begin{center}
  {\bf abstract}
\end{center}
We study some general properties of coupled quantum systems.  We consider simple interactions between two copies of identical Hamiltonians such as the SYK model,
Pauli spin chains with random magnetic field and harmonic oscillators. Such couplings make the ground states close to the thermofield double states of the uncoupled Hamiltonians.
For the coupled SYK model, we push the numerical computation further 
towards the thermodynamic limit so that an extrapolation in the size of the system is possible. 
We find good agreement between the extrapolated numerical result and the analytic result in the large-$q$ limit.
We also consider the coupled gauged matrix model and vector model,
and argue that the deconfinement is associated with the loss of the entanglement,
similarly to the previous observation for the coupled SYK model.
The understanding of the microscopic mechanism of the confinement/deconfinement transition
enables us to estimate the quantum entanglement precisely, and
backs up the dual gravity interpretation which relates the deconfinement to the disappearance of the wormhole.
Our results demonstrate the importance of the entanglement between the color degrees of freedom in the emergence of the bulk geometry
from quantum field theory via holography.

\newpage
\tableofcontents

\section{Introduction}
Quantum entanglement provides us with various curious phenomena, 
such as the quantum teleportation \cite{quantum_teleportation}.
From the point of view of holography, there is an interesting connection between
quantum entanglement and wormhole \cite{Maldacena:2001kr,Maldacena:2013xja}.
Recently it has been proposed that, by coupling two quantum systems appropriately,
it is possible to engineer a traversable wormhole \cite{Gao:2016bin}.
This provides a dual gravitational description of the quantum teleportation \cite{Gao:2016bin,Maldacena:2017axo} and also helps deepen our understanding of the puzzle of information loss during the evaporation of black holes~\cite{Almheiri:2019hni,Almheiri:2019yqk,Penington:2019kki,Almheiri:2019qdq}.

A simple and explicit model that realizes the setting in this context is the ``coupled SYK model" constructed in~\cite{Maldacena:2018lmt}.
As we will review in Sec.~\ref{sec:coupled-SYK}, this model has several features
of interest both from quantum mechanical and gravitational points of view.
We would like to know whether results obtained within this model are generic, and if yes,
try to give prescriptions useful for theoretical considerations and experimental studies.
Therefore, we investigate analogous prescriptions for spin, free fermions and bosonic systems,
and study their properties. Furthermore we study coupled gauged systems with gauge singlet constraints in order to understand the connection with a wider class of models of quantum gravity on a firmer footing.

A key player in this context is the thermofield double  state (TFD). In order to define the TFD, we introduce two copies of the identical Hilbert space
--- the ``left" and ``right" Hilbert spaces, denoted by ${\cal H}_{\rm L}$ and ${\cal H}_{\rm R}$ ---
on which two copies of identical Hamiltonian, denoted by $\hat{H}_{\rm L}$ and $\hat{H}_{\rm R}$, act.
The thermofield double at temperature $T=\beta^{-1}$ is given by
\begin{eqnarray}
|{\rm TFD}; \beta\rangle
=
\sum_{E}e^{-\frac{1}{2}\beta E}|E\rangle_L\otimes |E\rangle_R^\ast\ .
\label{eq:defTFD}
\end{eqnarray}
Here the sum runs over all energy eigenstates. The symbol ${}^*$ denotes the complex conjugate.
The TFD state is a purification of the thermal state, in the sense that
the reduced density matrix
$\hat{\rho}_{\rm L} = {\rm Tr}_{\rm R} |{\rm TFD}; \beta\rangle\langle {\rm TFD}; \beta|$
is the same as the thermal density matrix of the left system.
When the quantum theory admits a dual gravity description, the TFD state is dual to the eternal black hole, which is a maximally entangled black hole connected by the Einstein-Rosen bridge
\cite{Maldacena:2001kr,Maldacena:2013xja}.
In general, the TFD state is not the ground state of $\hat{H}_{\rm L}\otimes\textbf{1}+\textbf{1}\otimes\hat{H}_{\rm R}$.
The Einstein-Rosen bridge in the gravity dual, namely the eternal black hole, is not traversable.

Alternatively, one can add a certain coupling between the two copies, 
\begin{eqnarray}
\hat{H}_{\rm coupled}
=
\hat{H}_{\rm L}\otimes\textbf{1}
+
\textbf{1}\otimes\hat{H}_{\rm R}
+
\hat{H}_{\rm int},
\label{eq:coupled_model}
\end{eqnarray}
such that the ground state of the coupled system
mimics the TFD state of the uncoupled theory.
Such systems provide us with useful setups for studying the TFD state both numerically and experimentally. For instance, the ground state can be studied more easily than the TFD using numerical techniques such as the Markov Chain Monte Carlo. Also, it appears possible to realize such system experimentally by engineering the coupled Hamiltonian, with for instance suggested protocols using quantum wires or graphene flakes bilayers~\cite{Lantagne:2019}. When the system is in the regime that allows a gravitational dual description, the coupled model admits an interpretation as an eternal traversable wormhole~\cite{Gao:2016bin,Maldacena:2017axo,Maldacena:2018lmt}.

In this work, we focus on this coupled system protocol~\cite{Maldacena:2018lmt}, but we note that other ways for approximating the TFD state have also been proposed,  e.g.~\cite{Cottrell:2018ash,Martyn:2018wli}. 
Our first goal is therefore to understand which kind of coupling $\hat{H}_{\rm int}$ can achieve this objective for a variety of theories. This will be the topic of Sec.~\ref{sec:coupled-SYK}, Sec.~\ref{sec:cuopled-spin-system}, Sec,~\ref{ferm_coupled} and Sec.~\ref{sec:coupled-harmonic-oscillators}. Our second goal, pursued in Sec.~\ref{sec:matrix-model}, Sec.~\ref{sec:vector_model} and Sec.~\ref{sec:geometric-interpretation}, is to further improve the understanding about the relation between the quantum entanglement and spacetime.
We will study how the loss of the entanglement and disappearance of the traversable wormhole are related. In the past, this problem has been studied for the SYK model \cite{Maldacena:2018lmt}.  We will consider gauge theories, for which precise calculations and simple gravity picture are available based on the knowledge about the confinement/deconfinement transition.\footnote{See~\cite{Klebanov:2007ws} for a relationship between confinement and entanglement in another context.} 

The detailed organization of this paper is as follows. In Sec.~\ref{sec:coupled-SYK}, we review the coupled SYK model, setting the strategy for finding
the prescriptions for other systems. We additionally present numerical results for larger systems than previously available (up to $N=52$ fermions for $q=4$), as well as for a different value $q=8$.
In Sec.~\ref{sec:cuopled-spin-system}, we propose a prescription for spin systems, and check the validity by numerically studying a model of coupled spin chains in various parameter regimes.
We then provide in Sec.~\ref{ferm_coupled} the solution for a simple case of coupled free fermions systems.
In Sec.~\ref{sec:coupled-harmonic-oscillators}, we study a similar prescription for the harmonic oscillators. 
We derive analytic expressions for the ground state, which are useful for later sections. 
We also perform a numerical calculation to demonstrate the loss of entanglement in the excited states. 
Technical details on the numerical computations performed in the above sections are given in the Appendices.
In Sec.~\ref{sec:matrix-model}, the prescription considered in Sec.~\ref{sec:coupled-harmonic-oscillators} is applied to matrix models. The gauged Gaussian matrix model is studied in Sec.~\ref{sec:Gauged-Gaussian-MM},
and it is explained how deconfinement and loss of entanglement are related.
In Sec.~\ref{sec:BMN-matrix-model}, we briefly comment on the interacting matrix models.
In Sec.~\ref{sec:vector_model}, we perform similar analyses on the O($N$) vector model.
The dual gravity interpretation is provided in Sec.~\ref{sec:geometric-interpretation}.
The notion of partial deconfinement \cite{Hanada:2016pwv,Hanada:2018zxn,Hanada:2019czd,Hanada:2019kue,Berenstein:2018lrm}
enables us to estimate the quantum entanglement precisely, and 
backs up the dual gravity interpretation which relates deconfinement to the disappearance of a traversable wormhole.
\section{Coupled SYK Model}\label{sec:coupled-SYK}
As a concrete example of a coupled Hamiltonian \eqref{eq:coupled_model}, 
we consider the coupled SYK models \cite{Maldacena:2018lmt} by taking 

\begin{eqnarray}
\hat{H}_{\alpha}=(i)^{q/2} \sum_{1<j_1 <j_2<...<j_q}^{N/2} J_{j_1 \ldots j_q} 
\hat{\chi}_{\alpha}^{j_1} 
\hat{\chi}_{\alpha}^{j_2}
\ldots 
\hat{\chi}_{\alpha}^{j_q}
\end{eqnarray}
with $\alpha={\rm L}$ or ${\rm R}$ and 
\begin{eqnarray}
\hat{H}_{\rm int}
= 
i \mu \sum_{j=1}^{N/2} 
\hat{\chi}_{\rm L}^j 
\hat{\chi}_{\rm R}^j. 
\end{eqnarray}
Here $\hat{\chi}^i_{\rm L}$ and $\hat{\chi}^i_{\rm R}$ are Majorana fermion operators acting on the left and right copies, respectively, which satisfy the anticommutation relation 
$\{\hat{\chi}^i_\alpha,\hat{\chi}^j_\beta\}=\delta^{ij}\delta_{\alpha\beta}$. The random couplings $J_{j_1 \ldots j_q}$ are taken uniformly from a normal distribution of zero mean and variance 
$\left\langle J^2_{j_1 \ldots j_q}\right\rangle=\frac{2^{q-1} {\cal J}^2 (q-1)!}{q(N/2)^{q-1}}$.  We further set ${\cal J}=1$ as an energy scale. We used similar notations and normalizations as in~\cite{Maldacena:2018lmt} and~\cite{Garcia-Garcia:2019poj} (the later uses the notation $k$ for the coupling instead of $\mu$). There are $\frac{N}{2}$ fermions on each copy; $N$ is the total number of fermions in the coupled model. 

As $\mu\to\infty$, the ground state approaches the Fock vacuum in terms of $\hat{c}_i$'s:
\begin{eqnarray}
|{\rm GS}(\mu)\rangle
\to
|I\rangle
\qquad
(\mu\to\infty),  
\end{eqnarray}
where
\begin{eqnarray}
\hat{c}_i|I\rangle
=
0. 
\end{eqnarray} 
The maximally entangled state $|I\rangle$ is a TFD state at $T=\infty$. 

At finite $\mu$, $|{\rm GS}(\mu)\rangle$ is close to the TFD state at a certain temperature $\beta(\mu)$, which we denote by $|{\rm TFD},\beta(\mu)\rangle$~\cite{Maldacena:2018lmt}.  The inverse temperature $\beta(\mu)$ is determined to maximize the overlap ${\cal O}=| \langle {\rm TFD} (\beta) | {\rm GS}(\mu) \rangle|$ for each given $\mu$.

We reproduce previously published overlap results~\cite{Garcia-Garcia:2019poj,Lantagne:2019} for $q=4$ and extend them to larger systems, and also present new results for $q=8$. First, we compute the ground state of the coupled Hamiltonian \eqref{eq:coupled_model} and then compare it with the TFD state~\eqref{eq:defTFD}, which is obtained either directly from eqn.~\eqref{eq:defTFD} or computed by $|{\rm TFD},\beta\rangle \propto e^{-\frac{\beta}{4} ( \hat{H}_{L} + \hat{H}_{R})} |I\rangle$. We are able to push the calculations up to $N=52$ Majorana fermions for $q=4$ and $N=40$ for $q=8$ (see Appendix~\ref{sec:appendix_numerics} for details on the numerics).

\begin{figure}[htbp]
	\center{
		\includegraphics[width=0.45\columnwidth]{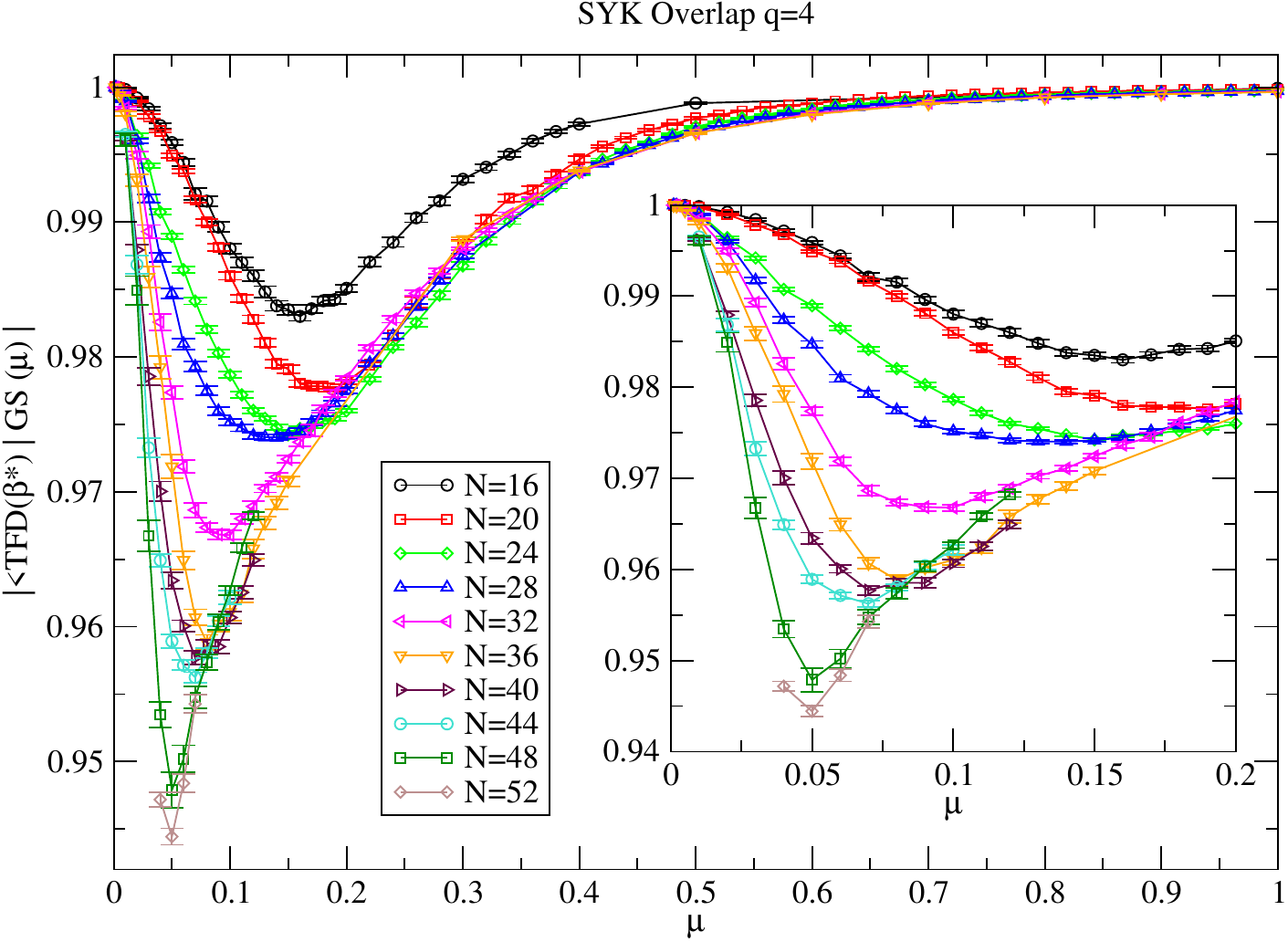}
		\includegraphics[width=0.45\columnwidth]{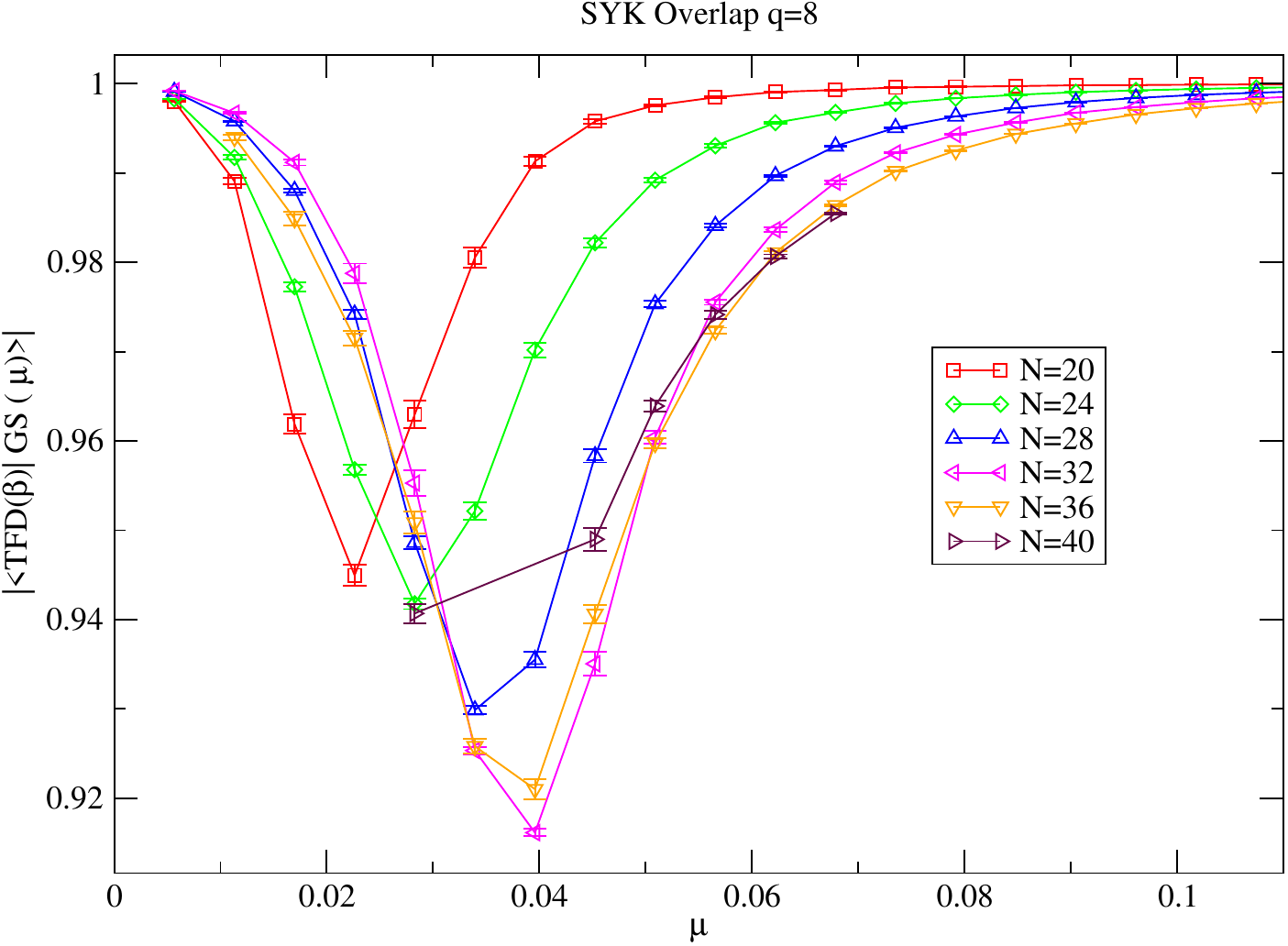}
	}
	\caption{\label{fig:overlap_SYK} Average overlap $\overline{{\cal O}}=\overline{| \langle {\rm TFD} (\beta) | {\rm GS}(\mu) \rangle|}$ at the optimal temperature, $
		\beta=\beta^\ast(\mu)$, for the $q=4$ (left panel) and $q=8$ (right panel) coupled SYK models, as a function of coupling strength $\mu$, for different system sizes $N$. Each data point is obtained as an average over at least 50 realizations of disorder. The inset in the left panel zooms on the region where the overlap is minimal.}
\end{figure}

The results for the average overlap $\overline{{\cal O}}$ are presented in Fig.~\ref{fig:overlap_SYK}. As expected, the overlap reaches unity in the two limiting cases of $\mu \rightarrow 0$ and $\mu \rightarrow \infty$, and differs from this value increasingly with system size for intermediate values of $\mu$. We expect the overlap to vanish in the thermodynamic limit for any non-zero finite $\mu$, even though the numerical results are not enough to conclude this. 
However it is quite remarkable that for systems with very large Hilbert spaces,  the TFD at the optimal $\beta$ differs only by a very small amount from the ground state of the coupled model: at most $5.5 \%$   for $q=4$ (up to ${\rm dim}{\cal H}=2^{26}\simeq 6.7\times 10^7$)  and $\sim 8 \%$ for $q=8$. 

Another interesting feature in Fig.~\ref{fig:overlap_SYK} is that the overlap displays a minimum for a coupling strength $\mu^*$ which moves considerably with system size, in particular for $q=4$. While for $N=28$ and $32$, we have $\mu^* \simeq 0.15,0.1$ for $q=4$ (as observed in previous work~\cite{Garcia-Garcia:2019poj,Lantagne:2019}), we obtain $\mu^* \simeq 0.05$ for $N=48,52$. The largest sizes for $q=8$ develop a minimum around a similar value $\mu^* \simeq 0.04$. It is hard based on our data to extrapolate to the thermodynamic limit behavior, in particular for $q=8$ where the minimum $\mu^*$ does not vary monotonously. Based solely on the $q=4$ data, it is possible that for larger $N$, $\mu^*$ continues to decrease together with a decreasing minimum overlap. 

\begin{figure}[h]
	\centering
	\includegraphics[width=0.7\linewidth]{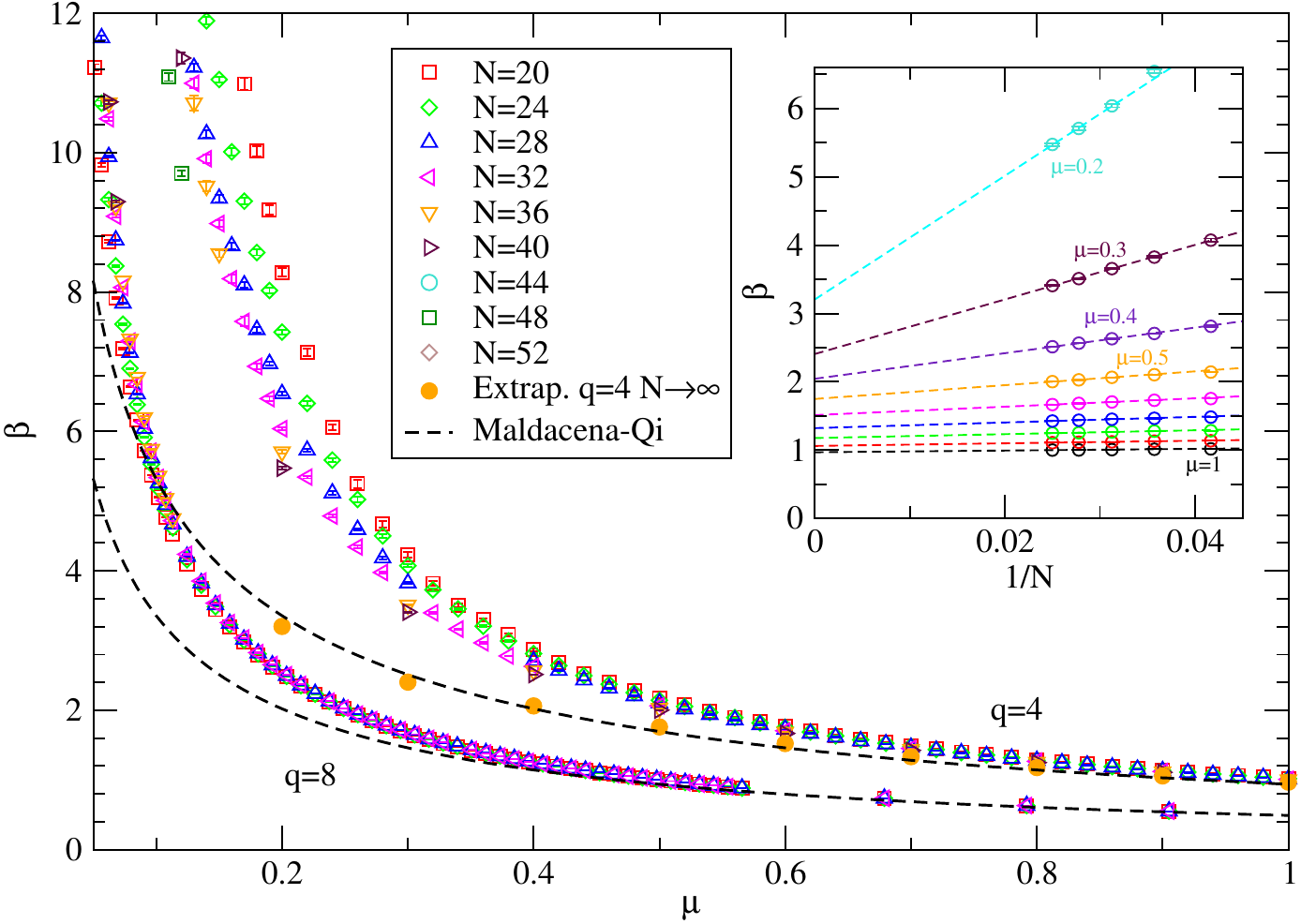}
	\caption{Effective inverse temperature as a function of ${\mu}$: comparison between the finite-$N$ data for $q=4$ and $q=8$ and the analytical expression from~\cite{Maldacena:2018lmt}. The inset shows the $1/N$ extrapolation for selected values of $\mu$ ($\mu=0.2,0.3,...,1$ from top to bottom): the dotted lines are linear fits over the last three points ($N=32,36,40$).}
	\label{fig:mubeta}
\end{figure}

We now compare our numerical results of the inverse temperature $\beta(\mu)$ to the analytic solution of the overlap ${\cal O}=| \langle {\rm TFD} (\beta) | {\rm GS}(\mu) \rangle|$ in the large-$q$ limit~\cite{Maldacena:2018lmt}. As shown in~\cite{Maldacena:2018lmt}, in the large-$q$ limit one can show that the overlap saturates $\left|\langle {\rm TFD} (\beta) | {\rm GS}(\mu) \rangle\right| =1$ for any $\mu$ at an effective inverse temperature $\beta(\mu)$ given by 
\begin{equation}
\beta(\mu)=\frac{2}{\alpha} \sqrt{1+\left(\frac{\alpha}{\mathcal{J}}\right)^{2}} \arctan \frac{\mathcal{J}}{\alpha},
\label{betamu}
\end{equation}
where  $\alpha=\mathcal{J} \sinh \gamma, \mu q=2\alpha \tanh \gamma,  \epsilon=\frac{{\mu}q}{2 \mathcal{J}}$. As we adopted the same normalization convention as in~\cite{Maldacena:2018lmt}, we can directly compare our results (taking ${\cal J}=1$) as presented in Fig.~\ref{fig:mubeta}. 
The agreement is very good when $\mu$ is sufficiently large, both for $q=4$ and $q=8$. We find a larger discrepancy in the small-$\mu$ region, which is below roughly the same scale where the overlap starts to decrease. For $q=4$ and for a range $0.2\lesssim \mu$, we attribute this to finite-$N$ effects. Indeed in this range, we can extrapolate results to the thermodynamic limit using a $1/N$ fit for large-enough $N$ (see inset of Fig.~\ref{fig:mubeta}), and the extrapolated values are in good agreement with the analytical expression. For lower values of $\mu$, we cannot extrapolate correctly the data due to the larger error bars caused by the greater fluctuations of $\beta$ from sample to sample. This is also the region where the overlap starts to deviate much more significantly from unity, and therefore we expect the agreement to be less good. In fact, it is also possible to identify the deviation from unit overlap with the possible $1/q$ corrections that one has to take into account in the large-$q$ analysis~\cite{Maldacena:2018lmt}. It is argued in~\cite{Maldacena:2018lmt} that such a deviation is due to the possible excitations due to the stronger left-right coupling to the coupled model so that it is not accurate to mimic the ground state of the coupled system as the TFD state at some temperature.\footnote{We hope to provide some further insights into the deviation of the overlap between the numeric and the large-$q$ analysis results in the future~\cite{PY}. We also notice the numerical analysis of a related model in~\cite{Nosaka:2019tcx}.}

We finally note that the finite-$N$ corrections are more complex to interpret in the $q=8$ case. We observe that in the same range of values of $\mu$ where we could extrapolate the $q=4$ data, the dependence of $\beta(\mu)$ on $N$ is not always monotonous: for the very small set of data that we have access currently, $\beta(\mu)$ tends to first increase with $N$ (in the {\it opposite} direction than the one expected from the analytical prediction), and then it tends to decrease again for larger $N$. Note that this effect is hardly visible on the scale of Fig.~\ref{fig:mubeta}.  We thus conclude that, despite the fact that the analytical expressions should be closer to the numerical data for large $q$, the range of $N$ that we can reach with $q=8$ is too small to be in an asymptotic regime where finite-$N$ effects are easily interpreted.

\section{Coupled Spin System}\label{sec:cuopled-spin-system}
Let us next consider Hamiltonians consisting of Pauli spin operators. For concreteness, we will use spin chains later, but for the moment, details of the interactions within the left or right systems are not needed to describe the coupling between them. Our strategy is simply to mimic the coupled SYK model by making sure that the TFD state at infinite temperature
$\sum_{E}|E\rangle_L\otimes |E\rangle_R^\ast=\frac{1}{\sqrt{2^N}}\left(|\!\!\uparrow\rangle|\!\!\uparrow\rangle^\ast+|\!\!\downarrow\rangle|\!\!\downarrow\rangle^\ast\right)^{\otimes N}$
becomes the ground state of the coupled Hamiltonian when the coupling is large. For this, we consider the following coupling:\footnote{
	We use the same notation $L$ for the left chain and the number of spins per chain to stick with conventional notations, the meaning of this label should be clear for all situations. 
}
\begin{eqnarray}
\hat{H}_{\rm int}
=
\mu
\sum_{i=1}^{L/2}\left(
\hat{\Sigma}_i^\dagger\hat{\Sigma}_i
+
\hat{\Sigma}_i\hat{\Sigma}^\dagger_i
\right),
\end{eqnarray}
\begin{eqnarray}
\hat{\Sigma}_i
=
\sigma_{iL}^+-(\sigma_{iR}^-)^\ast,
\qquad
\hat{\Sigma}^\dagger_i
=
\sigma_{iL}^--(\sigma_{iR}^+)^\ast,
\qquad  \sigma_{i\alpha}^\pm=\frac{\sigma_{i\alpha}^x\pm\sqrt{-1}\sigma_{i\alpha}^y}{2}
\end{eqnarray}
with $\alpha= L,R$. Since
$\hat{\Sigma}\left(|\!\!\uparrow\rangle|\!\!\uparrow\rangle^\ast+|\!\!\downarrow\rangle|\!\!\downarrow\rangle^\ast\right)
=\hat{\Sigma}^\dagger\left(|\!\!\uparrow\rangle|\!\!\uparrow\rangle^\ast+|\!\!\downarrow\rangle|\!\!\downarrow\rangle^\ast\right)=0$, the ground state at $\mu=\infty$ obtained from $\langle {\rm GS }|\hat{H}_{\rm int}| {\rm GS}\rangle=0$,
is $|{\rm GS }\rangle=\frac{1}{2^L}\left(|\!\!\uparrow\rangle|\!\!\uparrow\rangle^\ast+|\!\!\downarrow\rangle|\!\!\downarrow\rangle^\ast\right)^{\otimes L/2}$ indeed.

We turn to numerical experiments to check whether this simple type of coupling plays a similar role as in the SYK model. We study spin chains with random field:
\begin{equation}\label{Hspin}
\hat{H}_\alpha
=
\sum_{i=1}^{L/2}\left(
\frac{1}{4}
\vec{\sigma}_{i,\alpha}\vec{\sigma}_{i+1,\alpha}
+
\frac{\vec{w}_{i,\alpha}}{2}\vec{\sigma}_{i,\alpha}
\right),
\end{equation}
where the random magnetic field is chosen along the $z$-direction $\vec{w}=(0,0,w)$ and uniformly random in $[-W,+W]$ and $\alpha=L,R$ denotes the left / right chain. This model offers the opportunity to probe the adequateness of the TFD to mimic the ground state of different phases of coupled matter. For $W=0$, the model (Heisenberg model) is integrable. For $0 \leq W <W_c$, the system is in an chaotic/ergodic phase, while for $W > W_c$, the system is in a Many-Body Localized (MBL) phase. Current estimate of the critical disorder (in the middle of the spectrum) is~\cite{Luitz:2015} $W_c \simeq 3.7$. It is important to remark that the nature of the ground state of the {\it coupled} system does not necessarily reflect the underlying phase of the uncoupled system. This is indeed the case for the disorder-free model ($W=0$) where the coupled system is expected to be a gapped paramagnet for any $\mu>0$~\cite{Dagotto:1996}, whereas the ground-state of the uncoupled system is a critical liquid. While we are not aware of specific predictions for $W>0$, we expect a similar behavior (gapped paramagnetic ground state) for all systems in the limit $\mu \gg 1$. 

We study the coupled model Eq.~\eqref{eq:coupled_model} with the left $\hat{H}_L$ and right $\hat{H}_R$ systems taken with the same disorder realization $\{\vec{w}_i\}$ in Eq.~\eqref{Hspin}. 
We measure the average overlap $\overline{{\cal O}}$ between the ground state of the coupled system and the TFD and additionally compute the Kullback-Leibler (KL) divergence $ {\rm KL}  ={\rm Tr} \hat{\rho}_{L}\log\hat{\rho}_{L} - \hat{\rho}_{L}\log\hat{\sigma}(T)$,
where $\hat{\rho}_{L}={\rm Tr}_{\rm R}\left(|{\rm GS }\rangle\langle{\rm GS }|\right)$ is the reduced density matrix and $\hat{\sigma}(T)$ is the thermal density matrix of the uncoupled theory. 
The KL divergence is another measure of how different the TFD and the coupled ground state are: it vanishes in case they are equal.
We compute both the  inverse temperature $\beta({\cal O})$, which maximizes the overlap, and $\beta({\rm KL})$, which minimizes the KL-divergence. 

\begin{figure}[htbp]
	\begin{center}
		\includegraphics[width=0.75\columnwidth]{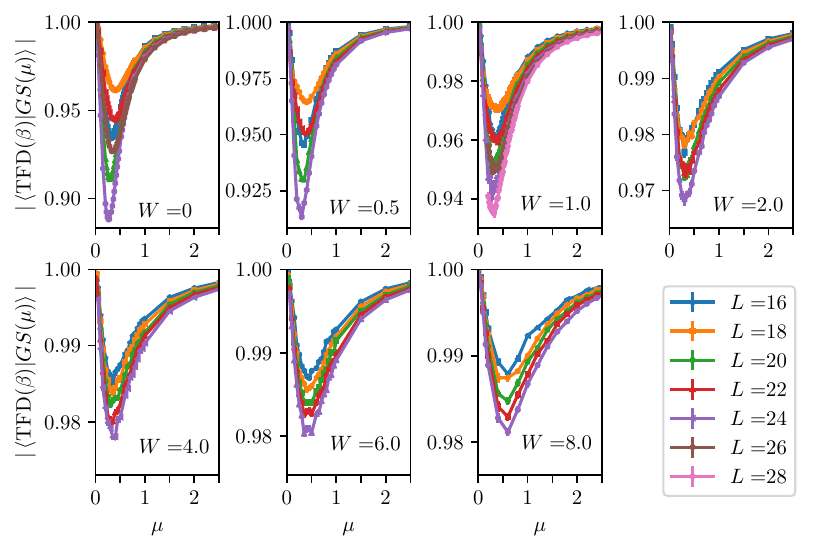}
	\end{center}
	\caption{Average overlap $\overline{{\cal O}}=\overline{| \langle {\rm TFD} (\beta) | {\rm GS}(\mu) \rangle|}$ at the optimal inverse temperature $\beta=\beta(\mu)$ as a function of coupling $\mu$ for the coupled spin chain model, for different strenghts of disorder $W$ and sizes $L$. Results are averaged over more than $100$ disorder realizations.
}\label{Fig:TFD-vs-gs1}
\vspace{-0.3cm}
\end{figure}

\begin{figure}[htbp]
	\begin{center}
		\includegraphics[width=0.48\columnwidth]{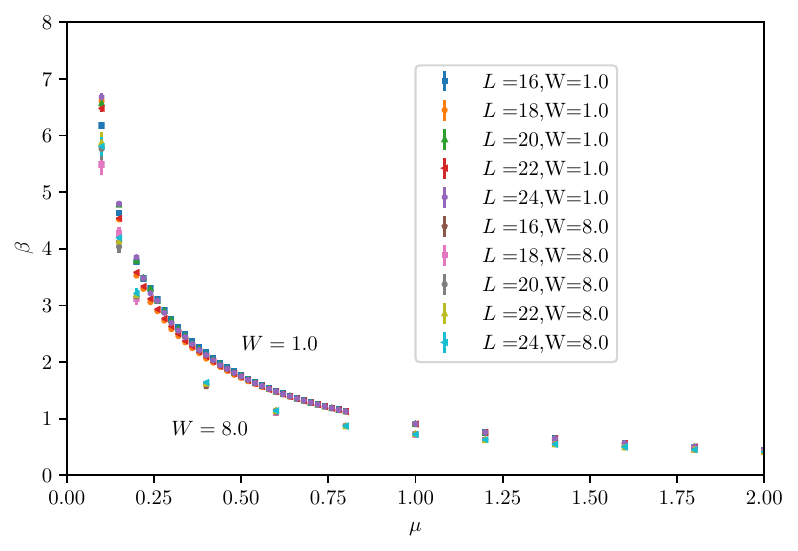}
		\includegraphics[width=0.48\columnwidth]{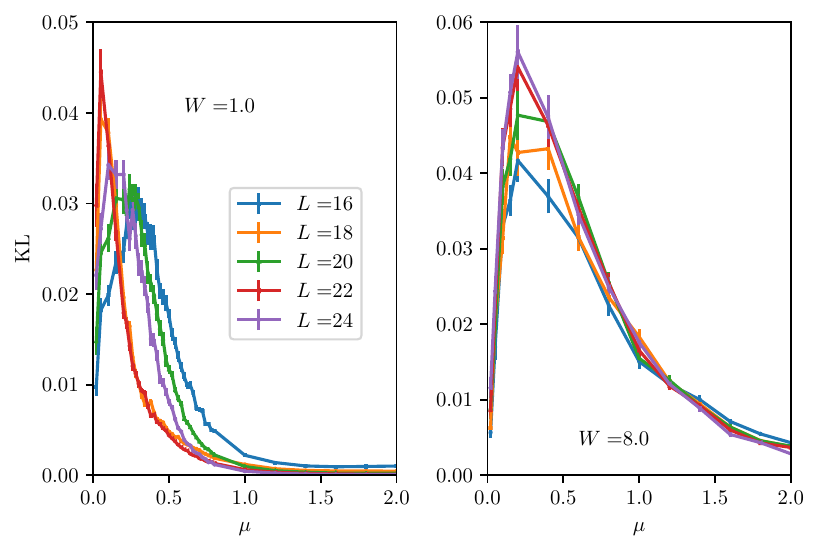}
		\includegraphics[width=0.6\columnwidth]{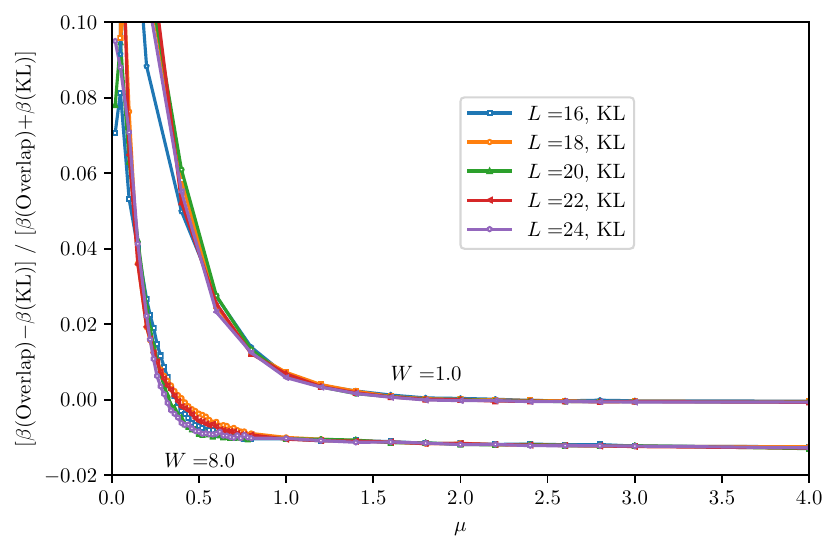}
	\end{center}
	\caption{Various quantities as a function of coupling $\mu$, for two disorder values ($W=1$ in the ergodic phase, $W=8$ in the Many-Body Localized phase) and for different sizes $L$, averaged over more than $100$ realizations of disorder. [Top, left] The optimal inverse temperature $\beta=\beta(\mu)$ which maximizes the overlap. [Top, right] The KL-divergence. [Bottom] Normalized difference between the inverse temperatures $\beta({\rm overlap})$ and $\beta({\rm KL})$ as determined from the maximization of the overlap or the minimization of the KL divergence.}\label{Fig:TFD-vs-gs2}
	\vspace{-0.3cm}

\end{figure}

Our numerics for the overlap (Fig.~\ref{Fig:TFD-vs-gs1}) show the same qualitative tendency, irrespective of whether the uncoupled system is located in the ergodic, MBL phases (small and large $W$, respectively) or integrable ($W=0$). Data are very similar to those of the coupled SYK model: the overlap is close to $1$ in both limits $\mu \rightarrow 0$ and $\mu \gg 1$, and displays a deviation from unity which increases with system size (with a non-trivial even-odd $L/2$ effect for small disorder strengths). This is corroborated by considering the small values of the KL divergence both in the ergodic and MBL cases of the uncoupled system: see bottom panel of Fig.\ref{Fig:TFD-vs-gs2}, where data for disorder $W=1,8$ are presented. The KL is maximal in the range where the overlap is minimal, as naturally expected, but its scale is overall quite small, considering the sizes of the vectors involved. The left top panel of Fig.\ref{Fig:TFD-vs-gs2} presents the inverse temperature $\beta$ as a function of coupling strength for these two values of disorder: again the data is very similar to the SYK case, even though there is no analytical prediction to compare with in this case (we also expect that $\beta$ scales as $1/\mu$ in the large $\mu$ limit, based on dimensional counting). We close these numerical observations by showing in the right top panel of Fig.\ref{Fig:TFD-vs-gs2} that the inverse temperatures as determined either by minimizing the KL divergence or the maximization of the overlap are very similar: the relative difference is at most a few percents, and non-negligible difference appears only in the very low $\mu$ regime where the error bars on $\beta$ are larger.

The conclusion of this numerical study of the coupled spin chains system is that the results appear very similar to the phenomenology obtained for the SYK model. In particular, one can find the inverse temperature $\beta(\mu)$ such that the ground state of the coupled model is very close to the TFD, at any value of the coupling strength $\mu$. Moreover, this appears to be the case irrespective of the underlying nature of the uncoupled system. 
We rationalize this with the following argument: when forming the TFD, a thermal bath with the explicit inverse temperature $\beta$ is imposed to the underlying doubled system. The sum over all energy levels (weighted by the Boltzmann factor) can completely wash out the individual eigenstate features. For instance, measuring the expectation value of a spin located, say, in the left system $\sigma^z_{i,L}$ in the TFD state gives the thermal expectation value (which vanishes in the $\beta \rightarrow 0$ limit), whereas it is well known that in the MBL phase, measuring expectation values over different eigenstates results in strongly different (polarized) values, even for eigenstates located in the middle of the spectrum (corresponding to $\beta  \rightarrow 0$)~\cite{MBLReview1,MBLReview2}. In other words, imposing the TFD completely washes out the localization features of the underlying MBL eigenstates, in the same way as a thermal bath destroys MBL; whether the uncoupled system is ergodic or localized does not play an important role here.  
Given that the features of the uncoupled theory --- whether it is ergodic or localized --- is irrelevant once the TFD state is formed, 
it would be reasonable to expect that such features does not play crucial roles in the formation of the TFD state.

\section{Coupled fermionic model}\label{ferm_coupled}

We  also consider coupled free fermions, in which case the simplest setting is to define the theory directly from the creation/annihilation operators. 

The thermofield double state at finite temperature of the fermionic system can be constructed as~\cite{Takahashi:1996zn}
\begin{align}
	|\text{TFD}\rangle = \frac{1}{\sqrt{1+e^{-\beta \omega} }}\left(1+e^{-\frac{\beta \omega}{2}}b_L^\dagger b_R^\dagger\right) |0,0\rangle\,,\label{thF}
\end{align}
where
\begin{align}
	\{b_L,b_L^\dagger\}=1\,,
	\qquad 
	\{b_R,b_R^\dagger\}=1\ .
\end{align}
The state~Eq.\eqref{thF} has the property that tracing over, say, the right degrees of freedom leads to the reduced density matrix  
\begin{align}
	\rho_L=\frac{1}{1+e^{-\beta \omega}}\left(|0\rangle\langle 0|+e^{-\beta \omega}|1\rangle\langle 1|\right)\,,	
\end{align}
that is a thermal density matrix.

To understand if it can be considered as a ground state of a coupled model, we consider the following general coupled free-fermionic model:
\begin{align}
	H=\omega b_L^\dagger b_L+\omega b_R^\dagger b_R +\mu_{1} b_L^\dagger b_R^\dagger+\mu_{2} b_L^\dagger b_R+\mu_{3} b_L b_R^\dagger+\mu_{4} b_L b_R+G\ .\label{Hall}
\end{align}  
Hermiticity requires 
\begin{align}
	\mu_4 = -\mu_1^*\,, \qquad \mu_2 = -\mu_3^*\,,
\end{align} 
Then 
\begin{align}
	H|TFD\rangle &=\left[\left(\frac{2\omega e^{-\beta \omega/2}}{\sqrt{1+e^{-\beta \omega}}}+\frac{\mu_1}{\sqrt{1+e^{-\beta \omega}}}\right)b_L^\dagger b_R^\dagger- \frac{\mu_4 e^{-\beta \omega/2}}{\sqrt{1+e^{-\beta \omega}}} \right]|0,0\rangle \propto |TFD\rangle\,,\label{eigH2}
\end{align}   
leads to
\begin{align}
	\left(\frac{2\omega e^{-\beta \omega/2}}{\sqrt{1+e^{-\beta \omega}}}+\frac{\mu_1}{\sqrt{1+e^{-\beta \omega}}}\right) \frac{1}{\sqrt{1+e^{-\beta \omega}}} = -\mu_4 \left(\frac{e^{-\beta \omega/2}}{\sqrt{1+e^{-\beta \omega}}}\right)^2 \ .
\end{align}
We can solve this equation to get the value $\mu_1$. If we further assume $\mu_1$ to be real, the result looks like
\begin{align}
	\mu_1 = -\frac{2 \omega  e^{\beta  \omega /2}}{e^{ \beta  \omega }-1}\ .
\end{align} 
Further notice that this condition does not impose any condition on the parameter $\mu_2$ and $\mu_3$. We simply set them to zero since they can simply be absorbed into a field redefinition that mixes $b_L$ with $b_R$.
This is effectively the relation between the coupling constants and the temperature of the  effective TFD state, analogous to ~Eq.\eqref{Teff} for the coupled oscillators (see next section). 

To further check that~Eq.\eqref{thF} is the ground state of the above Hamiltonian, we notice that the TFD state is annihilated by the following operators
\begin{align}
	f_L&=\frac{1}{\sqrt{1+e^{-\beta \omega}}}\left(b_L - e^{-\beta \omega/2 } b^\dagger_R\right)\\
	f_R&=\frac{1}{\sqrt{1+e^{-\beta \omega}}}\left(b_R + e^{-\beta \omega/2 } b^\dagger_L\right)\,,
\end{align}  
namely
\begin{align}
	f_L|\text{TFD}\rangle =0 =f_R|\text{TFD}\rangle\ .\label{lr0}
\end{align}
In terms of these operators, we find
\begin{align}
	&f_L^\dagger f_L+f_R^\dagger f_R\\
	& =\frac{1}{{1+e^{-\beta \omega}}}\left(\left(1 - e^{-\beta \omega } \right)b^\dagger_L b_L+\left(1 - e^{-\beta \omega } \right)b^\dagger_R b_R+2e^{-\beta \omega /2} b_L b_R-2e^{-\beta \omega /2} b_L^\dagger b_R^\dagger+2e^{-\beta \omega}\right)\ .
\end{align}
Thus we can rewrite~\eqref{Hall} into
\begin{align}
	H=\omega \frac{{1+e^{-\beta \omega}}}{1 - e^{-\beta \omega } } \left(f_L^\dagger f_L+f_R^\dagger f_R\right)+G-\frac{2 \omega  e^{-\beta \omega}}{1-e^{-\beta \omega}}\ .
\end{align} 
From this rewriting of the Hamiltonian, we conclude that its ground state should be annihilated by $f_L$ and $f_R$. This is nothing but the TFD state that we have defined in~Eq.\eqref{thF}.

Furthermore, the variable $G$ can be obtained by requiring the state to be a ground state with zero energy
\begin{align}
	H|\text{TFD}\rangle=0\,,
\end{align}  
which determines 
\begin{align}
	G= \frac{2 \omega}{e^{\beta  \omega }-1} =\frac{\mu_1^2}{\omega+\sqrt{\mu_1^2+\omega^2}}\ .
\end{align}

This then justifies our assertion that the state defined in~\eqref{thF} is a ground state of the coupled model with the Hamiltonian 
\begin{align}
	H=\omega b_L^\dagger b_L+\omega b_R^\dagger b_R +\m b_L^\dagger b_R^\dagger-\m b_L b_R+\frac{\mu^2}{\omega+\sqrt{\mu^2+\omega^2}}\ .\label{Hall2}
\end{align}  
The ground state of this Hamiltonian  is~\eqref{thF} where the inverse temperature $\beta$ is  determined from
\begin{align}
	\mu=-\frac{2 \omega  e^{\beta  \omega /2}}{e^{ \beta  \omega }-1}\ .
\end{align} 
This state is also a TFD state of the uncoupled $b_L$ and $b_R$ system with the TFD Hamiltonian
\begin{align}
	H_{\text{TFD}}=\omega \left(b^\dagger_L b_L-b^\dagger_R b_R\right)\,,
\end{align}
that satisfies
\begin{align}
	H_{\text{TFD}} |TFD\rangle =0\ .
\end{align}

\section{Coupled bosonic systems and confinement/deconfinement transition}

In the previous sections, we dealt with models with finite local Hilbert spaces (fermions and spins $1/2$). In this section, we instead consider models with bosonic degrees of freedom, which makes an exact unbiased numerical analysis practically impossible since these systems have infinite-dimensional Hilbert space.  One of our motivations to study bosonic systems is the application to high energy physics and quantum gravity, for which the gauge symmetry plays an important role. Gauge theories exhibit nontrivial phenomena, most notably the confinement/deconfinement transition, even in the weak-coupling limit which can be solved analytically to some extent~\cite{Sundborg:1999ue,Aharony:2003sx}. 

We start with the simplest but very instructive example, i.e. the coupled harmonic oscillators \cite{Srednicki:1993im}, in Sec.~\ref{sec:coupled-harmonic-oscillators}.
This system admits a numerical analysis with reasonable amount of computer resources, as long as we focus on the low-energy states.
We then use the results on the coupled harmonic oscillators to discuss the gauged matrix model in Sec.~\ref{sec:matrix-model}, and how the loss of the entanglement is related to the deconfinement. In Sec.~\ref{sec:vector_model}, we show that the same argument can be applied to the vector model. 
The confinement/deconfinement transition is related to the formation of black hole via gauge/gravity duality. 
In Sec.~\ref{sec:geometric-interpretation}, we finally discuss how our findings can be interpreted in the dual gravity picture.

\subsection{Coupled Harmonic Oscillators}\label{sec:coupled-harmonic-oscillators}
In order to construct coupled bosonic systems with interesting properties,
we start with the simplest but very instructive example, i.e. the coupled harmonic oscillator \cite{Srednicki:1993im}, with the following uncoupled Hamiltonian
\begin{eqnarray}
\hat{H}=\frac{\hat{p}^2}{2}+\frac{\omega^2\hat{x}^2}{2}\ .
\label{def-harmonic-oscillator}
\end{eqnarray}
Following the strategy adopted for the SYK, spin chain and free fermions models, 
we introduce a coupling between two identical copies of the harmonic oscillators such that the ground state of the coupled Hamiltonian
mimics the TFD state at infinite temperature.
The coupling we introduced is parameterized by two independent coupling constants $C_1$ and $C_2$:
\begin{eqnarray}
\hat{H}
=
\frac{\hat{p}_{\rm L}^2}{2}+\frac{\omega^2\hat{x}_{\rm L}^2}{2}
+
\frac{\hat{p}_{\rm R}^2}{2}+\frac{\omega^2\hat{x}_{\rm R}^2}{2}
-
C_1
\left(
\hat{x}_{\rm L}^2
+
\hat{x}_{\rm R}^2
\right)
+
C_2\left(
\hat{x}_{\rm L}
-
\hat{x}_{\rm R}
\right)^2\ .
\end{eqnarray}

The physical intuition that suggests a direct connection between the ground state of this model and a thermofield double state is the following. We first set $C_2$ to zero and vary $C_1$, the coupled model is equivalent to a pair of uncoupled oscillators with frequency $\sqrt{\omega^2-2C_1}$.
As $C_1$ approaches $\frac{\omega^2}{2}$, the wave function spreads out,
and the ground state approaches to
$\int_{-\infty}^\infty dx_{\rm L}\int_{-\infty}^\infty dx_{\rm R}|x_{\rm L}\rangle|x_{\rm R}\rangle$.
Next we turn on $C_2$ to a large value. Then $x_{\rm L}$ and $x_{\rm R}$ are forced to become close,
and hence the normalized ground state of the coupled system becomes close to
$\int_{-\infty}^\infty dx|x\rangle|x\rangle$. This is nothing but the TFD state at infinite temperature.
In the rest of this section we provide a quantitative analysis that verifies such intuition explicitly. 

For the convenience of later analysis, we reparametrize the deformation terms such that
\begin{eqnarray}
\hat{H}
=
\frac{\hat{p}_{\rm L}^2}{2}+\frac{\omega^2\hat{x}_{\rm L}^2}{2}
+
\frac{\hat{p}_{\rm R}^2}{2}+\frac{\omega^2\hat{x}_{\rm R}^2}{2}
+
\frac{C_+}{2}\left(
\hat{x}_{\rm L}
+
\hat{x}_{\rm R}
\right)^2
-
\frac{C_-}{2}\left(
\hat{x}_{\rm L}
-
\hat{x}_{\rm R}
\right)^2\ .
\label{eq:coupled_oscillators}
\end{eqnarray}
Notice that in principle we could separate out the $\frac{C_+-C_-}{2}\left(\hat{x}_{\rm L}^2+\hat{x}_{\rm R}^2\right)$ piece in the left-right coupling from the genuine ``interaction" term $\hat{x}_L\hat{x}_R$.
We choose not to do so because, as we will show below, the ground state of the coupled system is identical to a TFD state
if the coupling satisfies  $\sqrt{1+2C_+/\omega^2}=\left(\sqrt{1-2C_-/\omega^2}\right)^{-1}$.
It is just a matter of taste; if we interpret $\sqrt{\omega^2+\frac{C_+-C_-}{2}}$ to be the `original' frequency,
the ground state would then be the TFD state of the `shifted' frequency $\omega$.\footnote{
	As shown in~\cite{Srednicki:1993im}, the reduced density matrix can be regarded as the thermal density matrix
	with certain shifted frequency and effective temperature.
	If we take $C_+$ and $C_-$ such that $\sqrt{1+2C_+/\omega^2}=\left(\sqrt{1-2C_-/\omega^2}\right)^{-1}$,
	this shifted frequency agrees with $\omega$.
}
\subsubsection{Ground state as the TFD}
The Hamiltonian of the uncoupled theory is given by Eq.~\eqref{def-harmonic-oscillator}, 
with the canonical commutation relation $[\hat{x},\hat{p}]=i$.
The creation and annihilation operators are defined by 
%
$\hat{a}^\dagger=\sqrt{\frac{\omega}{2}}(\hat{x}-i\frac{\hat{p}}{\omega})$
and 
${a}=\sqrt{\frac{\omega}{2}}(\hat{x}+i\frac{\hat{p}}{\omega})$. The Hamiltonian can be written in terms of the number operator $\hat{n}=\hat{a}^\dagger\hat{a}$ as $\hat{H}=\omega\left(\hat{n}+\frac{1}{2}\right)$.
The vacuum $|0\rangle$ is defined by $\hat{a}|0\rangle=0$, and the normalized excited states are constructed as $|n\rangle=\frac{\hat{a}^{\dagger n}}{\sqrt{n!}}|0\rangle$.

The coupled model is defined by \eqref{eq:coupled_oscillators}, which can be rewritten as
\begin{eqnarray}
\hat{H}
=
\frac{\hat{p}_{\rm +}^2}{2}+\frac{\omega_+^2\hat{x}_{\rm +}^2}{2}
+
\frac{\hat{p}_{\rm -}^2}{2}+\frac{\omega_-^2\hat{x}_{\rm -}^2}{2}\,,\label{hcp}
\end{eqnarray}
where
\begin{eqnarray}
\omega_+=\sqrt{\omega^2+2C_+},
\qquad
\omega_-=\sqrt{\omega^2-2C_-},
\qquad
\hat{x}_\pm
=
\frac{\hat{x}_{\rm L}\pm\hat{x}_{\rm R}}{\sqrt{2}},
\qquad
\hat{p}_\pm
=
\frac{\hat{p}_{\rm L}\pm\hat{p}_{\rm R}}{\sqrt{2}}\ .
\end{eqnarray}
Note that both $\omega_+$ and $\omega_-$ are different from $\omega$ for any nonvanishing $C_\pm$.
The creation operators are
\begin{eqnarray}
\hat{a}^\dagger_\pm
=
\frac{r_\pm+r_\pm^{-1}}{2\sqrt{2}}\left(
\hat{a}_{\rm L}^\dagger
\pm
\hat{a}_{\rm R}^\dagger
\right)
-
\frac{r_\pm-r_\pm^{-1}}{2\sqrt{2}}\left(
\hat{a}_{\rm L}
\pm
\hat{a}_{\rm R}
\right),
\end{eqnarray}
where $r_\pm=\sqrt{\frac{\omega_\pm}{\omega}}$.
The ground state that satisfies $\hat{a}_+|0\rangle_{\rm coupled}=\hat{a}_-|0\rangle_{\rm coupled}=0$ is
\begin{equation}
|0\rangle_{\rm coupled}
=
{\cal N}^{-1/2}
e^{\frac{1}{4}\frac{r_+-r_+^{-1}}{r_++r_+^{-1}}(\hat{a}_{\rm L}^\dagger+\hat{a}_{\rm R}^\dagger)^2}
e^{\frac{1}{4}\frac{r_--r_-^{-1}}{r_-+r_-^{-1}}(\hat{a}_{\rm L}^\dagger-\hat{a}_{\rm R}^\dagger)^2}
|0\rangle_{\rm L}|0\rangle_{\rm R}
\label{coupled-oscillators-gs}
\end{equation}
where the normalization factor ${\cal N}$ is given by
\begin{eqnarray}
{\cal N}
=
\left(1-\left(\frac{r_+-r_+^{-1}}{r_++r_+^{-1}}\right)^2\right)^{-1/2}
\cdot
\left(1-\left(\frac{r_--r_-^{-1}}{r_-+r_-^{-1}}\right)^2\right)^{-1/2}
\label{zero-mode-coupled-normalization-H}
\end{eqnarray}

The ground state \eqref{coupled-oscillators-gs} can be rewritten as
\begin{eqnarray}
|0\rangle_{\rm coupled}
=
{\cal N}^{-1/2}
e^{A_1(\hat{a}_{\rm L}^{\dagger 2}+\hat{a}_{\rm R}^{\dagger 2})}
e^{A_2\hat{a}_{\rm L}^\dagger\hat{a}_{\rm R}^\dagger}
|0\rangle_{\rm L}|0\rangle_{\rm R}\,,
\end{eqnarray}
where
\begin{eqnarray}
A_1=\frac{1}{4}\frac{r_+-r_+^{-1}}{r_++r_+^{-1}}+\frac{1}{4}\frac{r_--r_-^{-1}}{r_-+r_-^{-1}},
\qquad
A_2=\frac{1}{2}\frac{r_+-r_+^{-1}}{r_++r_+^{-1}}-\frac{1}{2}\frac{r_--r_-^{-1}}{r_-+r_-^{-1}}.
\label{eq:A1_A2}
\end{eqnarray}
This is a TFD state when $A_1=0$, or equivalently 
\begin{equation}
\sqrt{1+2C_+/\omega^2}\sqrt{1-2C_-/\omega^2}=1, 
\qquad
\omega_+ \omega_-=\omega^2, 
\qquad
r_+ r_-=1. 
\label{rel1}
\end{equation}
With this condition, 
$A_2$ simplifies to the following expression:
\begin{eqnarray}
A_2=\frac{r_+-r_+^{-1}}{r_++r_+^{-1}}
=
\frac{\sqrt{1+\frac{2C_+}{\omega^2}}-1}{\sqrt{1+\frac{2C_+}{\omega^2}}+1}\ .
\end{eqnarray}
We can rewrite this ground state into a form that resembles the TFD state in a manifest manner
\begin{eqnarray}
|0\rangle_{\rm coupled}
&=&
{\cal N}^{-1/2}e^{A_2\hat{a}_{\rm L}^\dagger\hat{a}_{\rm R}^\dagger}
|0\rangle_{\rm L}|0\rangle_{\rm R}
\nonumber\\
&=&
{\cal N}^{-1/2}
\sum_n
A_2^n
|n\rangle_{\rm L}|n\rangle_{\rm R}
\nonumber\\
&=&
{\cal N}^{-1/2}|A_2|^{-1/2}
\sum_n
e^{-E_n/2T_{\rm eff}}
|n\rangle_{\rm L}|n\rangle'_{\rm R}\,,
\end{eqnarray}
where
\begin{eqnarray}
E_n=\left(n+\frac{1}{2}\right)\omega,
\qquad
|A_2|=e^{-\omega/2T_{\rm eff}}\ .
\end{eqnarray}
and
\begin{eqnarray}
|n\rangle'_{\rm R}
=
\left\{
\begin{array}{cc}
|n\rangle_{\rm R} & (A_2\ge 0) \\
(-)^n|n\rangle_{\rm R} & (A_2<0)
\end{array}
\right.\ .
\end{eqnarray}
From this expression we can read off the effective temperature for any given $C_{\pm}$ that satisfy~\eqref{rel1}
\begin{eqnarray}
T_{\rm eff}
=
-
\frac{\omega}{2\log\left(\left|\frac{r_+-r_+^{-1}}{r_++r_+^{-1}}\right|\right)}\ .\label{Teff}
\end{eqnarray}

By construction, $A_2$ can take  values between $-1$ and $+1$.
At $A_2=\pm 1$, $\omega_+$ or $\omega_-$ becomes zero, and the theory becomes ill-defined with a continuous spectrum; 
the effective temperature $T_{\rm eff}$ becomes infinite there. 
For all other values of $A_2$, we have shown that the ground state of the coupled model~\eqref{eq:coupled_oscillators} satisfying~\eqref{rel1} is identical to a TFD state at temperature~\eqref{Teff}. Such a TFD state is analogous to the one originally discussed in~\cite{Maldacena:2001kr}. 

Further notice that all the above discussion assumes $\omega_\pm^2>0$, and together with the condition~\eqref{rel1} they are the incarnation in this model of the fact that there is only one good choice of the sign of the coupling in order to make the wormhole traversable in the general discussion~\cite{Gao:2016bin}. 
In the coupled SYK model~\cite{Maldacena:2018lmt} there is a symmetry $\psi_L \leftrightarrow \psi_R$, $\mu \leftrightarrow -\mu$ so this effect is not observed. However in our coupled bosonic model since swapping left with the right copy is itself a symmetry and does not require to flip the sign of $\mu$, we expect to see such an effect. Our results indeed leads to the expected result (although this model might not have a pure gravity holographic dual). We thus think this is describing the same phenomenon discussed in~\cite{Gao:2016bin}.
\subsubsection{Decay of entanglement at high temperature}
In this section, we consider the excited states of the coupled model, instead of the ground state, where the coupling constants still satisfy  $\sqrt{1+2C_+/\omega^2}\sqrt{1-2C_-/\omega^2}=1$.
It is expected that the quantum entanglement between the two sides is washed away when sufficiently large amount of energy is added to the system.
We check this expectation quantitatively in this section.

We consider the excited state
\begin{equation}
|n_+,n_-\rangle_{\rm coupled}
=
\frac{\hat{a}_+^{\dagger n_+}\hat{a}_-^{\dagger n_-}}{\sqrt{n_+!n_-!}}
|0\rangle_{\rm coupled}\,,\label{excit}
\end{equation}
where the energy of the system is 
\begin{equation}
E_{n_+,n_-}
=
\left(
n_+
+
\frac{1}{2}
\right)\omega_+
+
\left(
n_-
+
\frac{1}{2}
\right)\omega_-\ .
\end{equation}
We would like to study how does the entanglement between the left and right sectors of the system, which is in the state~\eqref{excit}, decay at high temperature/high energy.
In the following, we carry out some of our computations numerically imposing a cutoff $\Lambda$ to the Hilbert space. For details of the numerical methods, see Appendix~\ref{sec:numerical_calculation_harmonic_oscillator}.
\subsubsection*{Overlap with TFD}
\begin{figure}[htbp]
	\begin{center}
		\rotatebox{-90}{
			\includegraphics[width=5.5cm]{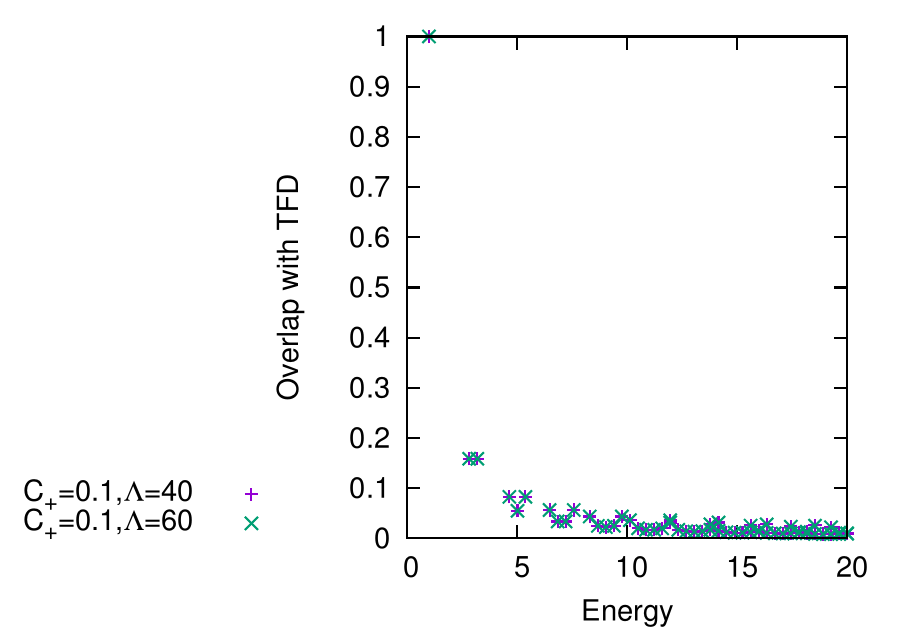}}
		\rotatebox{-90}{
			\includegraphics[width=5.5cm]{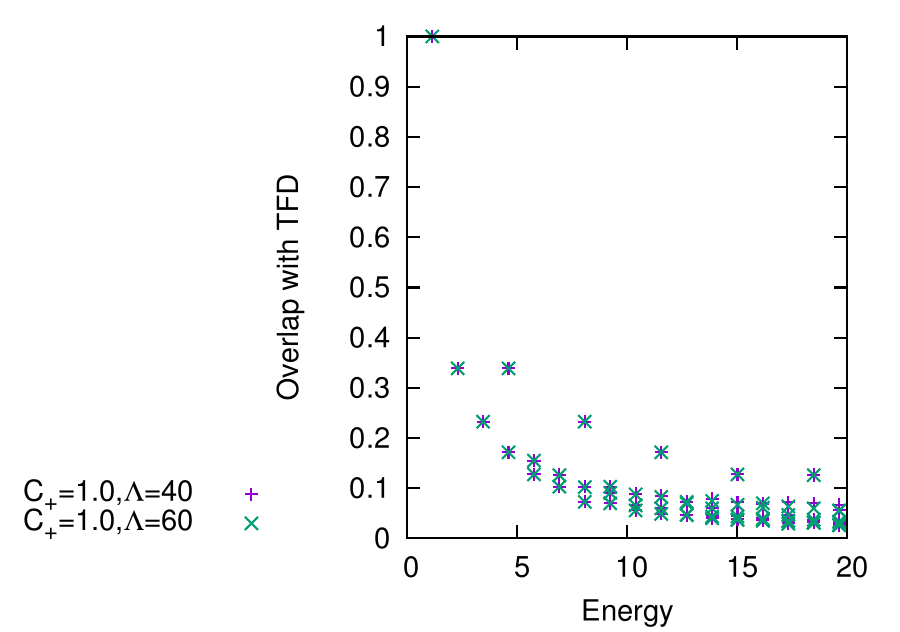}}
	\end{center}
	\caption{$|\langle {\rm TFD}(\beta)|n_+,n_-\rangle_{\rm coupled}|^2$ vs energy,
		for $C_+=0.1$ and $C_+=1.0$, 
		$\omega=1$. 
		At each energy, the value of $\beta$ is chosen 
		so that the overlap is maximized.
		Note that the overlap is exactly zero
		when $n_+$ or $n_-$ is odd (not shown in the plot).
	}\label{Fig:overlap_cp=010}
\end{figure}

We can calculate the overlap with the TFD state,
\begin{eqnarray}
\left|\langle {\rm TFD}(\beta)|n_+,n_-\rangle_{\rm coupled}\right|^2\ . 
\end{eqnarray}
For each $(n_+,n_-)$, the value of $\beta$ is chosen 
so that the overlap is maximized.
The results for $\omega=1$, $C_+=0.1$ and $1.0$ are shown in Fig.~\ref{Fig:overlap_cp=010}.
We can see that the overlap with TFD becomes smaller as the energy (or equivalently the temperature)
is increased. Note that the temperature of the TFD state
which maximizes the overlap depends on the given excited state, and except for the ground state this temperature is rather high.
Note also that $(n_+,n_-)=(a,b)$ and $(b,a)$ give the same overlap.\footnote{
	$(n_+,n_-)=(a,b)\mapsto(b,a)$ is realized by $\hat{a}_\pm\mapsto\mp\hat{a}_\mp$, 
	or equivalently, $\hat{a}_{\rm L}\mapsto -\hat{a}_{\rm R}$, $\hat{a}_{\rm R}\mapsto \hat{a}_{\rm L}$.
	If $|n_+=a,n_-=b\rangle=\sum_{mn}c_{mn}|m\rangle_{\rm L}|n\rangle_{\rm R}$, 
	then $|n_+=b,n_-=a\rangle=\sum_{mn}(-1)^mc_{mn}|n\rangle_{\rm L}|m\rangle_{\rm R}$.
	Either way, the overlap with the TFD is $(\sum_p e^{-\beta E_p} c_{pp}^2)/(\sum_p e^{-\beta E_p})$. 
}

\subsubsection*{Mutual Information}\label{sec:mutual_information}
With the general discussion~\cite{Gao:2016bin} and the particular example of the SYK model~\cite{Maldacena:2018lmt} in mind, we would like to relate the correlation of the two sides with the existence of possible wormhole phases. As known previously~\cite{Shenker:2013pqa,Marolf:2013dba,Balasubramanian:2014gla}, the left-right propagator is not alway a good diagnose of this connection. In the rest of this section, we consider a refined mutual information, namely $S_{\rm EE, L}
+
S_{\rm EE, R}
-
S_{\rm diag}$,   to probe the quantum correlation of the left and right sides and demonstrate that the connection between the coupling of the two sides and the quantum entanglement between them.

\begin{figure}[htbp]
	\begin{center}
		\rotatebox{-90}{
			\includegraphics[width=3.5cm]{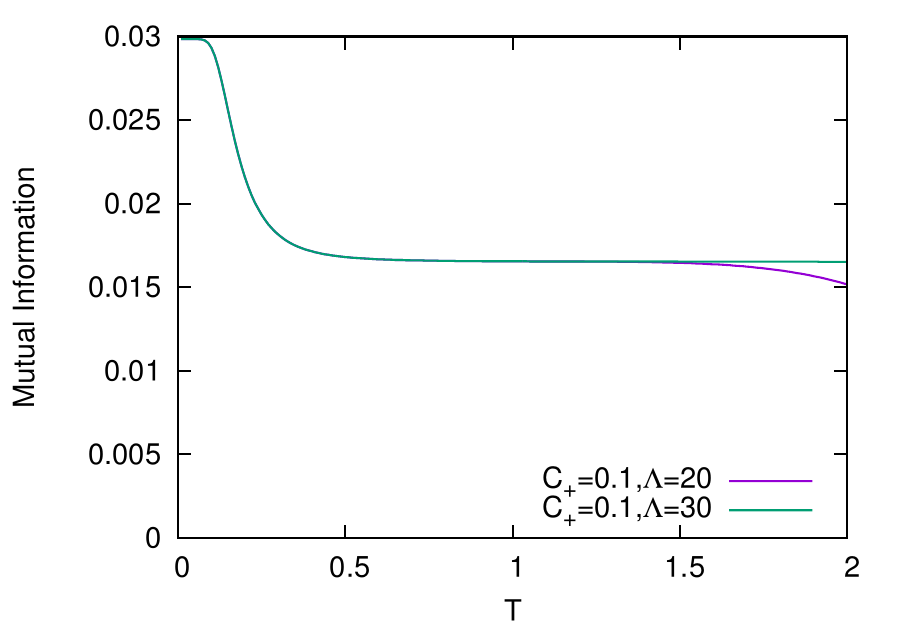}}
		\rotatebox{-90}{
			\includegraphics[width=3.5cm]{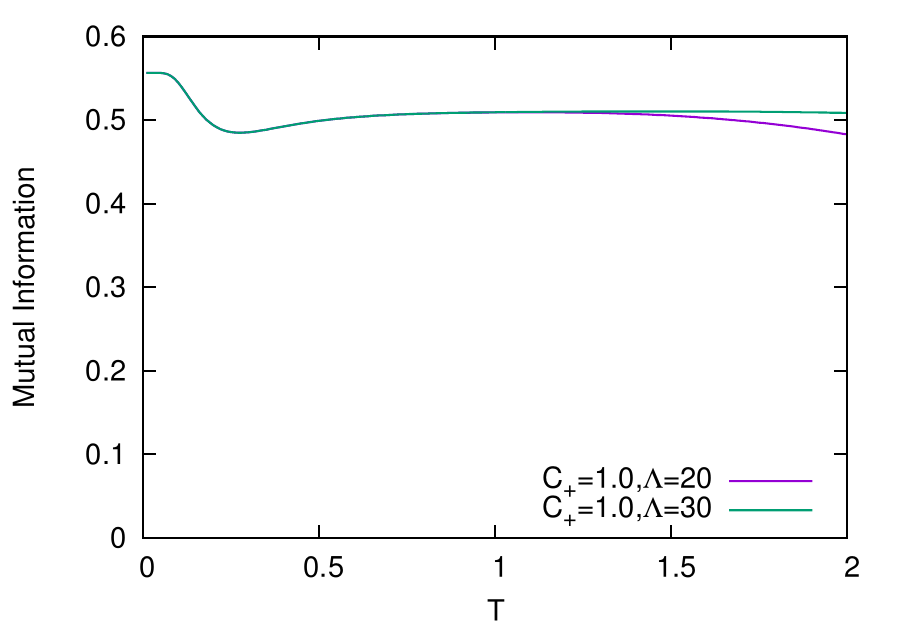}}
		\rotatebox{-90}{
			\includegraphics[width=3.5cm]{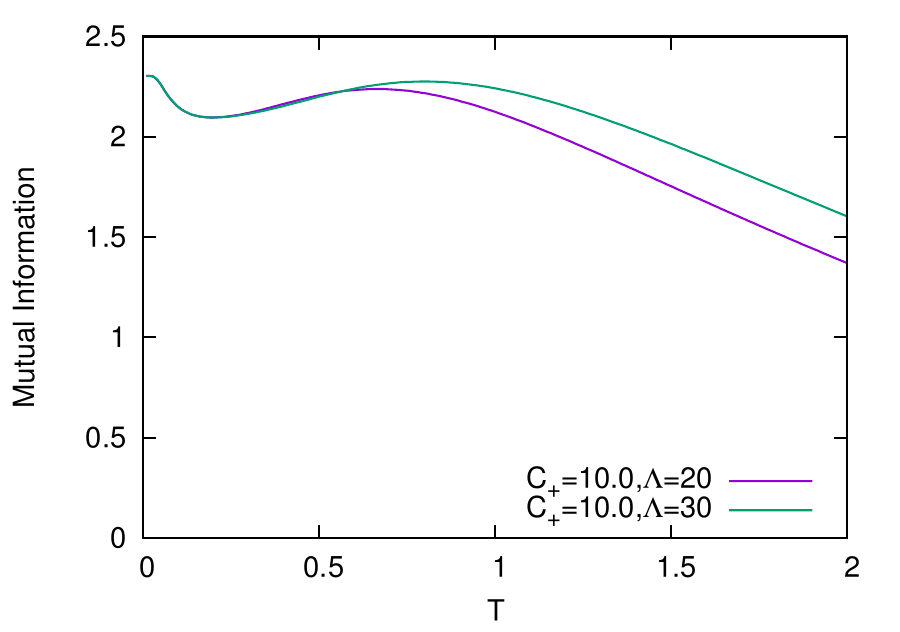}}
	\end{center}
	\caption{Mutual Information $S_{\rm EE, L}+S_{\rm EE, R}-S_{\rm therm}$ at $C_+=0.1, 1.0$ and $10.0$, $\omega=1$.
	}\label{Fig:MI-C01}
\end{figure}

The mutual information (MI) is defined by
\begin{eqnarray}
S_{\rm EE, L}
+
S_{\rm EE, R}
-
S_{\rm therm}\ .
\end{eqnarray}
Here $S_{\rm EE, L}$ and $S_{\rm EE, R}$ are the entanglement entropy obtained from 
the reduced density matrices $\hat{\rho}_{\rm L,R}={\rm Tr}_{\rm R,L}\hat{\rho}$, 
while $S_{\rm therm}$ is the thermal entropy of the entire system. 
As we can see from Fig.~\ref{Fig:MI-C01}, the MI does not decay significantly,
even when $C_+$ is as small as 0.1.
It is because the MI picks up both the classical and quantum correlations.
At high temperature, the MI should be dominated by the classical correlation.
On the other hand, the decrease of MI at low temperature reflects the decay of the quantum entanglement.
In order to study the quantum correlation more clearly, we need to subtract the classical correlation.

\begin{figure}[htbp]
	\begin{center}
		\rotatebox{-90}{
			\includegraphics[width=3.5cm]{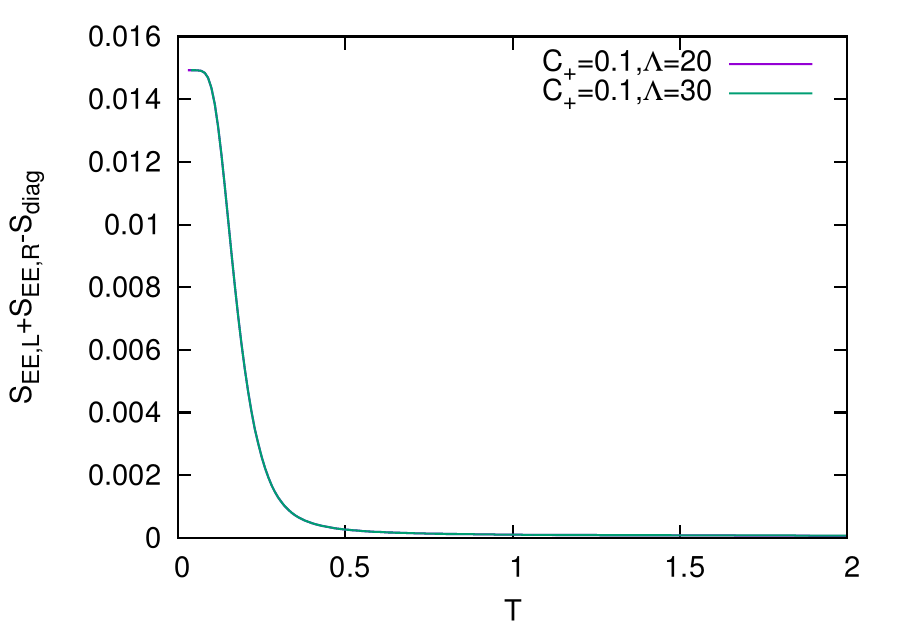}}
		\rotatebox{-90}{
			\includegraphics[width=3.5cm]{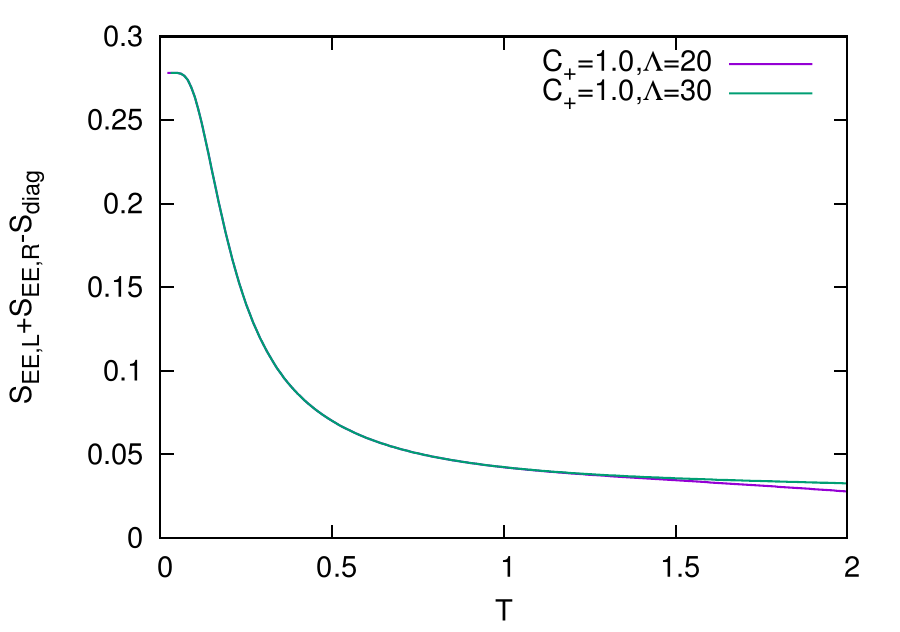}}
		\rotatebox{-90}{
			\includegraphics[width=3.5cm]{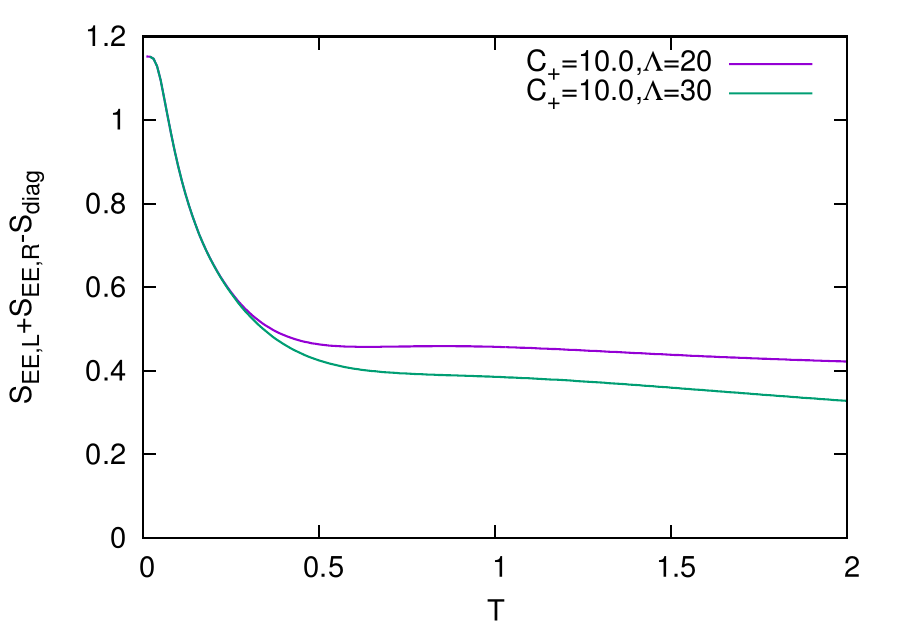}}
	\end{center}
	\caption{$S_{\rm EE, L}
		+
		S_{\rm EE, R}
		-
		S_{\rm diag}$, $C_+=0.1, 1.0$ and $10.0$, $\omega=1$, $\Lambda=20$ and $30$.
	}\label{Fig:quantum_correlation}
\end{figure}

To do so, recall that the thermal density matrix of the coupled system can be written in the following form:
\begin{eqnarray}
\hat{\rho}
=
\sum_{n_{\rm L},n_{\rm R},n'_{\rm L},n'_{\rm R}}
\rho_{n_{\rm L},n_{\rm R};n'_{\rm L},n'_{\rm R}}
\left(|n_{\rm L}\rangle\langle n'_{\rm L}|\right)
\left(|n_{\rm R}\rangle\langle n'_{\rm R}|\right)\ .
\end{eqnarray}
When the coupling parameters $C_+$ and $C_-$ are small,
tiny off-diagonal elements are generated and contribute to the entanglement.
If we keep only the diagonal part and define a separable state
\begin{eqnarray}
\hat{\rho}_{\rm diag}
=
\sum_{n_{\rm L},n_{\rm R}}
\rho_{n_{\rm L},n_{\rm R};n_{\rm L},n_{\rm R}}
\left(|n_{\rm L}\rangle\langle n_{\rm L}|\right)
\left(|n_{\rm R}\rangle\langle n_{\rm R}|\right)\,,
\end{eqnarray}
it should not capture the entanglement, rather it should capture
the classical thermal correlation between left and right sectors.
If $S_{\rm EE, L}+S_{\rm EE, R}-S_{\rm thermal}$ is not different from
$S_{\rm diag}-S_{\rm thermal}$, where $S_{\rm diag}=-{\rm Tr}\left(\hat{\rho}_{\rm diag}\log\hat{\rho}_{\rm diag}\right)$,
it is natural to expect that the MI is not picking up the quantum entanglement that is not due to thermal effects.
Therefore we use the quantity
\begin{equation}
S_{\rm EE, L}+S_{\rm EE, R}-S_{\rm diag}=S_{\rm EE, L}+S_{\rm EE, R}-S_{\rm thermal}-(S_{\rm diag}-S_{\rm thermal})\,,\label{smodify}
\end{equation} 
to characterize the quantum correlation between the left and right sides.
As we can see from Fig.~\ref{Fig:quantum_correlation}, the difference
$S_{\rm EE, L}
+
S_{\rm EE, R}
-
S_{\rm diag}$ decays significantly when $C_\pm$ are small.
We interpret it as the evidence that the quantum entanglement decays as temperature goes up.

Note that $S_{\rm diag}$ manifestly depends on the choice of the basis of the Hilbert space.
It is possible that a more elaborate choice of the basis could lead to a better estimate of the classical correlation. This is indeed aligned with the fact that there does not seem to be a canonical measure of  the classical contribution to general entanglement entropy. We consider the quantity~\eqref{smodify} because the entropy so defined monotonically decreases as temperature increases, as shown in Fig.~\ref{Fig:MI-C01}, which is what we expect since raising temperature generally destroys quantum entanglement. See e.g.~\cite{Chen:2019qqe} for discussions of entanglement entropy in the coupled SYK model.   

\subsubsection{The case of $C_+= C_-$}
\begin{figure}[htbp]
	\begin{center}
		\rotatebox{-90}{
			\includegraphics[width=3.5cm]{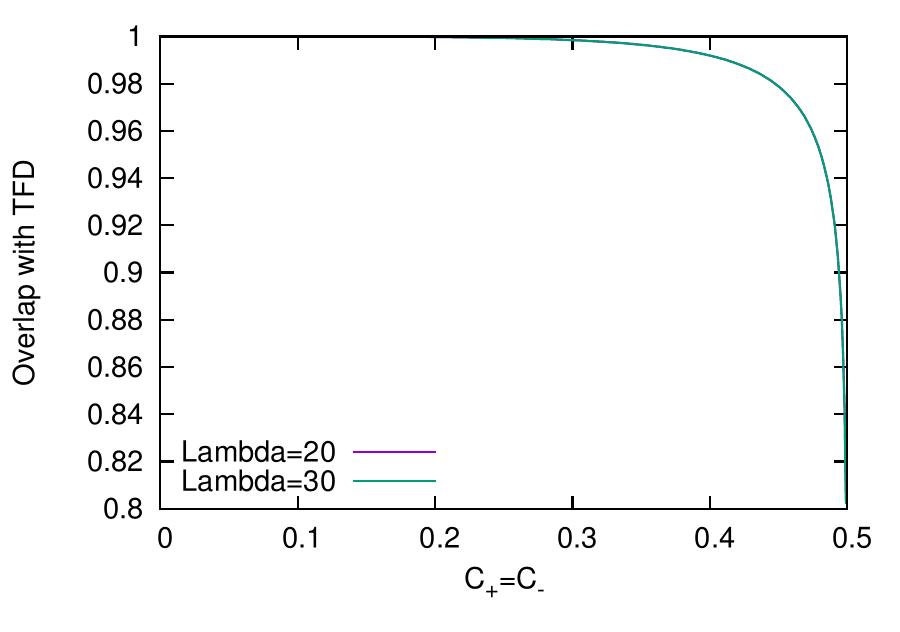}}
		\rotatebox{-90}{
			\includegraphics[width=3.5cm]{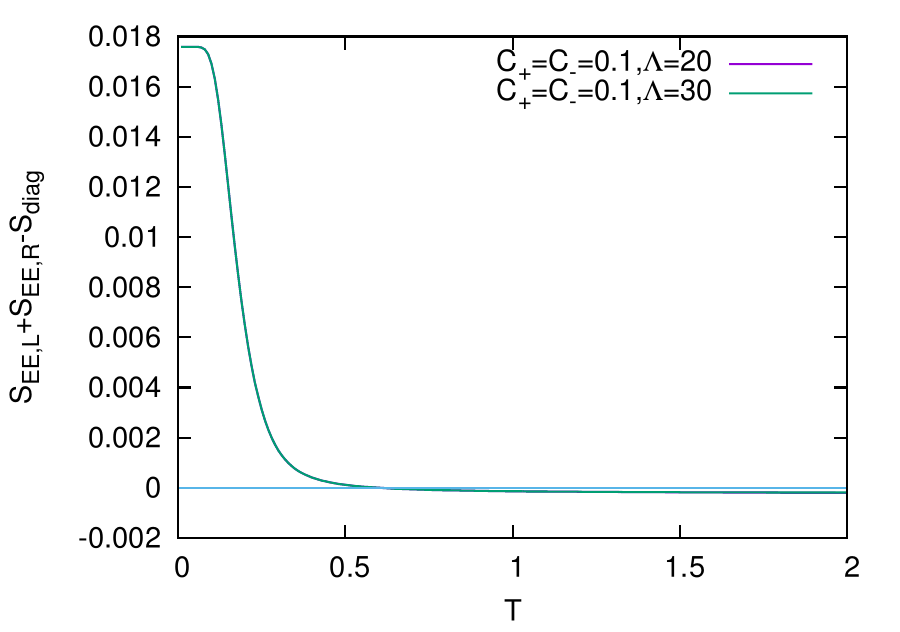}}
		\rotatebox{-90}{
			\includegraphics[width=3.5cm]{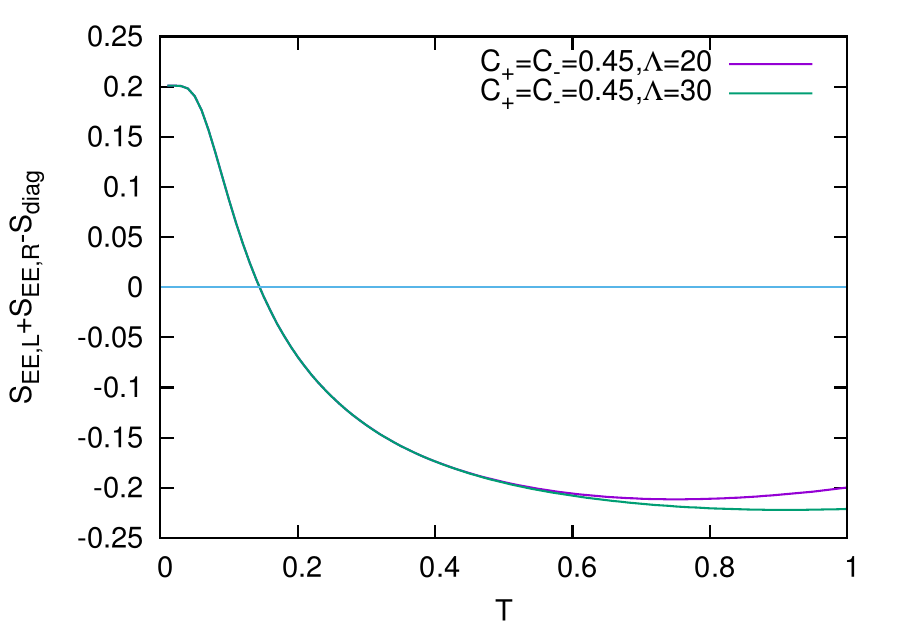}}
	\end{center}
	\caption{
		Overlap and $S_{\rm EE, L}
		+
		S_{\rm EE, R}
		-
		S_{\rm diag}$ for the case with $C_+=C_-=C$. The plots are for $C=0.1$ and $0.45$ respectively where $\omega=1$ and  $\Lambda=20$ and $30$ for each choice of $C$.
	}\label{Fig:cp=cm}
\end{figure}

$\hat{\rho}_{\rm diag}$, $S_{\rm diag}$ and the overlap with the TFD state explicitly depend on
the choice of `uncoupled' and `interaction' parts,
unlike the entanglement entropy and mutual information.
Therefore, let us consider another natural example:
\eqref{eq:coupled_oscillators} with $C_+=C_-=C$, namely
\begin{eqnarray}
\hat{H}
=
\frac{\hat{p}_{\rm L}^2}{2}+\frac{\omega^2\hat{x}_{\rm L}^2}{2}
+
\frac{\hat{p}_{\rm R}^2}{2}+\frac{\omega^2\hat{x}_{\rm R}^2}{2}
+
2C
\hat{x}_{\rm L}\hat{x}_{\rm R}.
\end{eqnarray}
In this case, the ground state of the coupled system is not the TFD state in the L-R basis, since we can check that~\eqref{rel1} is not satisfied in this case for all $C>0$. 
The system is well-defined when $\omega_\pm^2=\omega^2\pm 2C\ge 0$, namely $-\frac{\omega^2}{2}\le C\le\frac{\omega^2}{2}$.
As we can see from Fig.~\ref{Fig:cp=cm}, the ground state is close to the TFD as long as the coupling does not become too large,
and
$S_{\rm EE, L}
+
S_{\rm EE, R}
-
S_{\rm diag}$
is small.

\subsection{Coupled matrix models}\label{sec:matrix-model}
In this section, we apply the results obtained in Sec.~\ref{sec:coupled-harmonic-oscillators} to gauged matrix models,
which is closely related to physics of black holes via gauge/gravity duality.
The models exhibit a deconfinement transition:
at the critical temperature $T=T_c$, the energy of the system suddenly jumps a large amount. 
Below $T_c$, the energy and entropy are of order $N^0$ (after subtracting the zero-point energy); hence the $O(N^2)$ color degrees of freedom are `confined' and not visible individually. 
Above $T_c$, the energy and entropy are of order $N^2$; the color degrees of freedom are `deconfined' and become visible. 
The deconfinement transition corresponds to the formation of black hole in the gravity side
via holography. 

\subsubsection{Gauged Gaussian matrix model}\label{sec:Gauged-Gaussian-MM}
Let us consider the simplest, analytically calculable example: the gauged Gaussian matrix model.
The Euclidean action is given by
\begin{eqnarray}
S
=
N\sum_{I=1}^D\int_0^\beta dt {\rm Tr}
\left(
\frac{1}{2}(D_tX_I)^2
+
\frac{1}{2}X_I^2
\right).
\label{action:GMM-uncoupled}
\end{eqnarray}
The covariant derivative $D_tX_I$ is defined by $D_tX_I=\partial_tX_I-i[A_t,X_I]$,
where $A_t$ is the gauge field. When the gauge field is integrated out, the gauge-singlet constraint emerges.
Other than the singlet constraint, this system is nothing but
a bunch of non-interacting $DN^2$ harmonic oscillators with $m=\omega=1$.

The coupled version is
\begin{eqnarray}
S
&=&
N\int_0^{\beta} dt {\rm Tr}
\left(
\frac{1}{2}(D_tX_I)^2
+
\frac{1}{2}X_I^2
\right)
+
N\int_0^{\beta} dt {\rm Tr}
\left(
\frac{1}{2}(D_tY_I)^2
+
\frac{1}{2}Y_I^2
\right)
\nonumber\\
& &
\qquad
+
\frac{NC_+}{2}\int_0^{\beta} dt {\rm Tr}
\left(
X_I+Y_I
\right)^2
-
\frac{NC_-}{2}\int_0^{\beta} dt {\rm Tr}
\left(
X_I-Y_I
\right)^2.
\end{eqnarray}
We use the same gauge field for the left and right copies,
so that $X$ and $Y$ transform as the adjoints under the same SU($N$) gauge group,
because otherwise the coupling term is not gauge invariant.
Having the results in Sec.~\ref{sec:coupled-harmonic-oscillators} in mind,
let us take $C_-$ such that
\begin{eqnarray}
1+2C_+
=
\frac{1}{1-2C_-}
\end{eqnarray}
is satisfied. 
Then the ground state can be interpreted as a product of TFD's discussed in Sec.~\ref{sec:coupled-harmonic-oscillators}. 

\begin{figure}[htbp]
	\begin{center}
		\scalebox{0.12}{
			\includegraphics{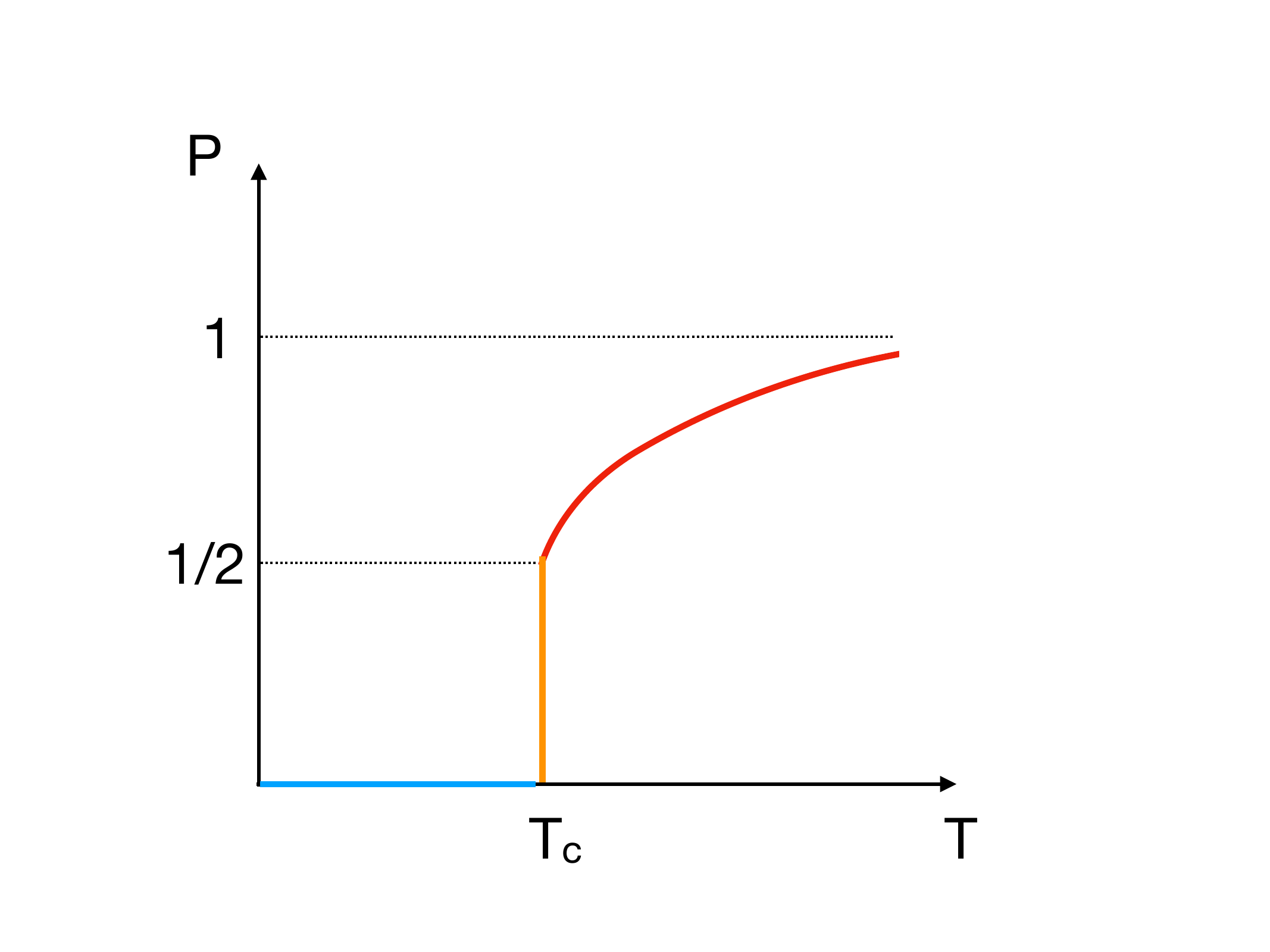}}
		\scalebox{0.12}{
			\includegraphics{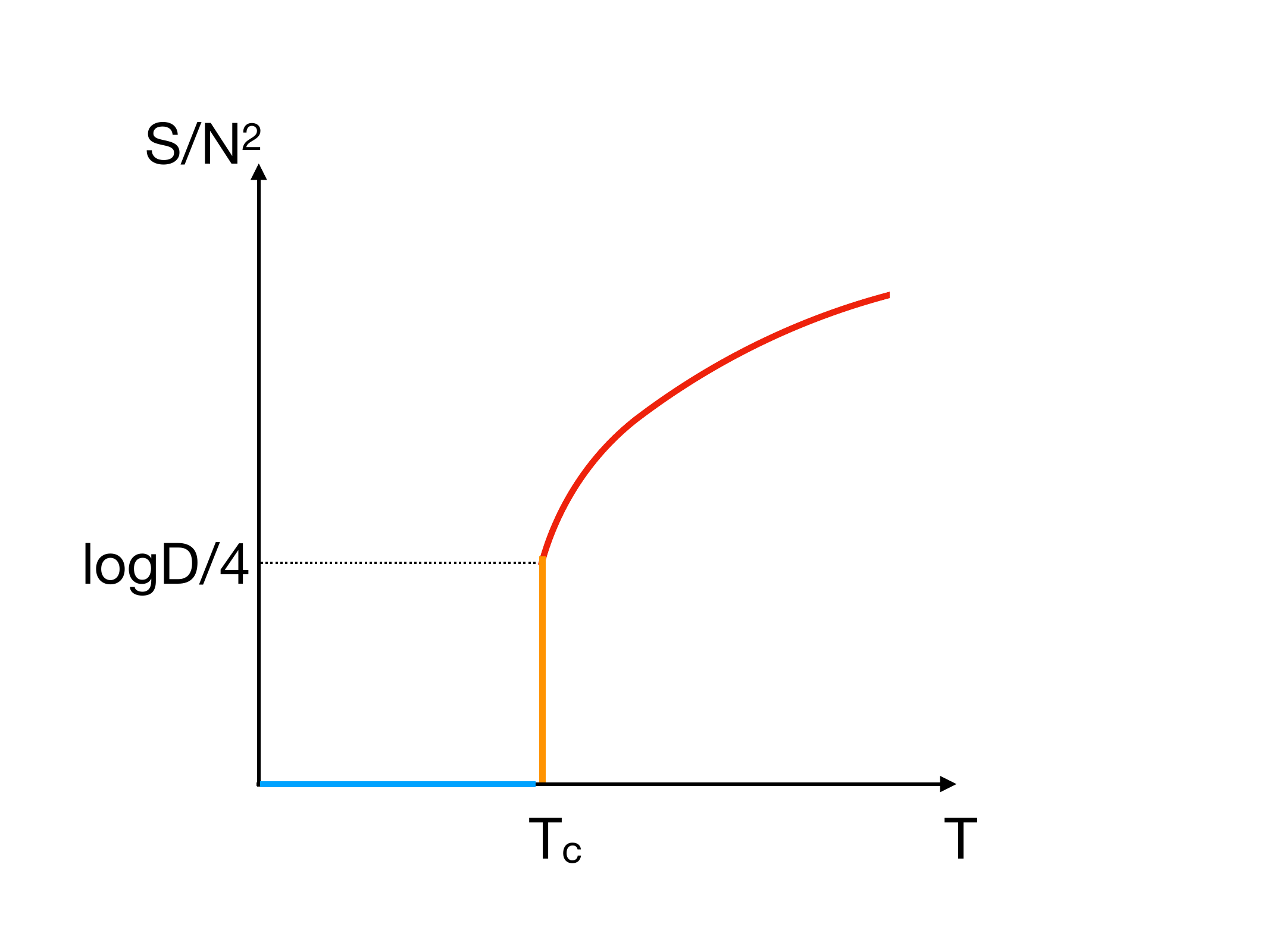}}
		\scalebox{0.12}{
			\includegraphics{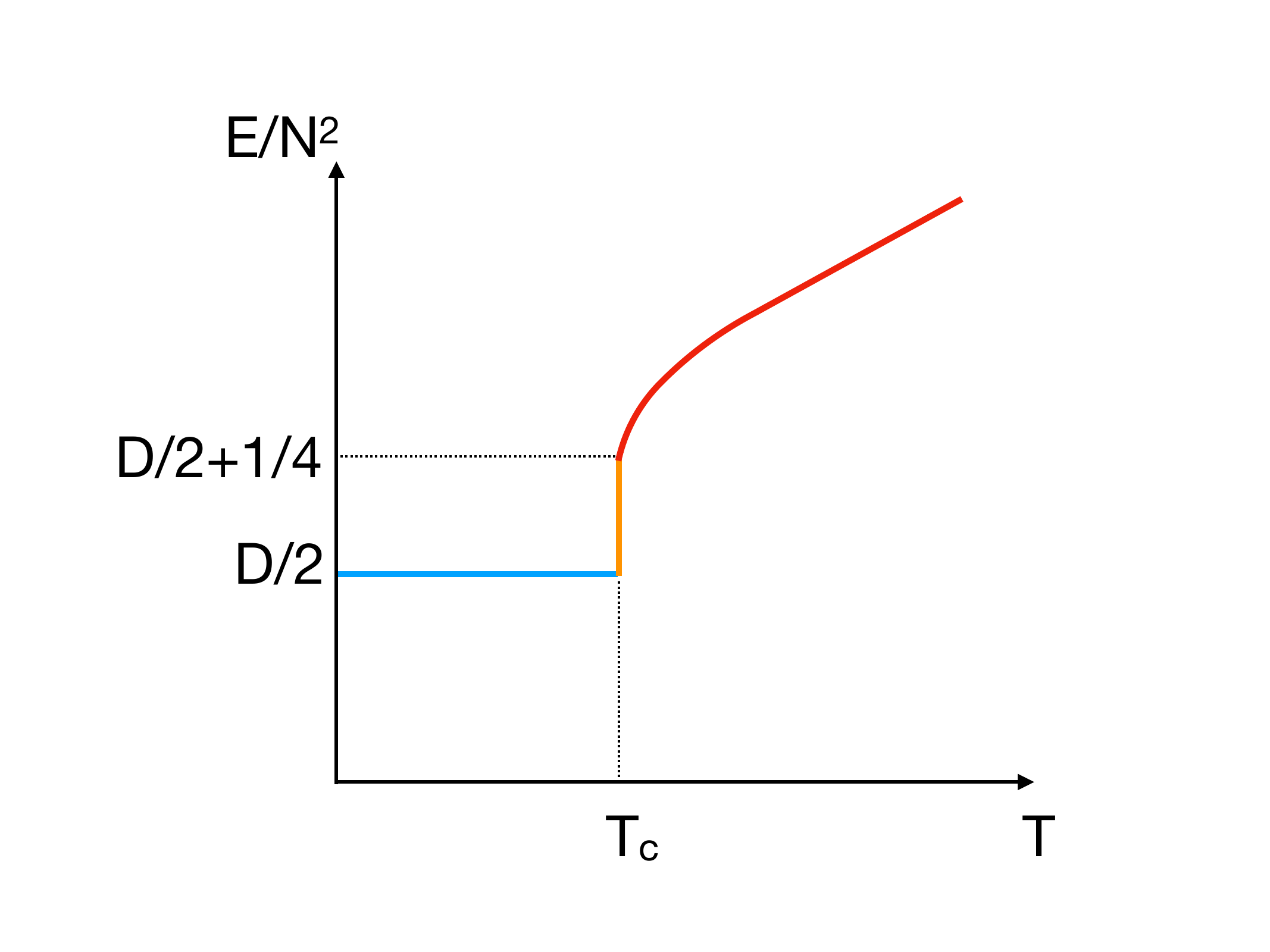}}
	\end{center}
	\caption{
		Sketches of the temperature dependence of the Polyakov loop $P$, entropy $S$ and energy $E$ in the gauged Gaussian matrix model \eqref{action:GMM-uncoupled}. 
		Blue, orange and red lines are identified with the confined, partially deconfined and completely deconfined phases, respectively. 
		These figures are taken from Ref.~\cite{Hanada:2019czd}. 
	}\label{fig:transition_GMM}
\end{figure}

The gauged Gaussian matrix model exhibits the confinement/deconfinement transition
because of the gauge-singlet constraint (see e.g.~\cite{Berenstein:2018lrm,Hanada:2019czd}).
The free energy of the original, single-copy uncoupled theory is\footnote{
	The analysis presented below is essentially the same as 4d Yang-Mills on S$^3$ \cite{Sundborg:1999ue,Aharony:2003sx}. 
}
\begin{eqnarray}
\beta F
&=&
\log Z(\beta)
\nonumber\\
&=&
\frac{N^2D}{2}
\log\left(\det\left(-D_t^2+1\right)\right)
-
\frac{N^2}{2}
\log\left(\det\left(-D_t^2\right)\right)
\nonumber\\
&=&
\frac{DN^2\beta}{2}
+
N^2\sum_{n=1}^\infty
\frac{1-Dx^n}{n}|u_n|^2,
\end{eqnarray}
where $x=e^{-\beta}$, $u_n=\frac{1}{N}{\rm Tr}P^n$ and ${\cal P}={\rm diag}(e^{i\theta_1},\cdots,e^{i\theta_N})$ is the Polyakov line.
Strictly speaking, this is the effective action in terms of $\theta_1,\cdots,\theta_N$; the free energy is obtained by minimizing it with respect to $\theta$'s. 
The term 
$\frac{N^2D}{2}
\log\left(\det\left(-D_t^2+1\right)\right)$ is the contribution from $D$ scalars,
while $\frac{N^2}{2}
\log\left(\det\left(-D_t^2\right)\right)$ is associated with the gauge fixing.
There is a first order phase transition
at $T_c=\frac{1}{\log D}$, where $|u_1|$ jumps from 0 to $\frac{1}{2}$; 
see Fig.~\ref{fig:transition_GMM} 
(It cannot go beyond $\frac{1}{2}$ because the density distribution $\rho(\theta)=\frac{1}{2\pi}\left(1+2u_1\theta\right)$
must not be less than zero). 
This leads to the first order phase transition without hysteresis.
At $T>T_c$, $u_2, u_3,\cdots$ become nonzero as well, while $u_1$ becomes larger,
so that the free energy is minimized while $\rho(\theta)$ remains non-negative.

If we consider the microcanonical ensemble (i.e.~use the energy as a parameter, rather than the temperature),
a rich structure can be found at the phase transition.
The key concept is the partial deconfinement
\cite{Hanada:2016pwv,Hanada:2018zxn,Hanada:2019czd,Hanada:2019kue,Berenstein:2018lrm}:
an SU($M$) subgroup of SU($N$) deconfines, and $\frac{M}{N}$ increases from zero to one as energy grows,
as $E\propto M^2$. The value of $M$ is $M=2N|u_1|$ in this case. 
When the SU($M$) subgroup is deconfined, $N^2-M^2$ degrees of freedom remain confined, namely they remain as the ground state.

For the coupled model, we obtain
\begin{eqnarray}
\beta F =-\log Z(\beta)
=
\frac{DN^2\beta (\omega+\omega^{-1})}{2}
+
N^2\sum_{n=1}^\infty
\frac{1-D(x^n+x^{\prime n})}{n}|u_n|^2,
\end{eqnarray}
where $\omega=\sqrt{1+2C_+}$,  $x=e^{-\beta\omega}$ and $x'=e^{-\beta /\omega}$.

As temperature is raised, the deconfinement phase transition takes place
at $1-D(x+x')=0$. The critical temperature is
$T_c=\frac{1}{\log 2D}$ for $C_+=0$ and $T_c=0$ for $C_+=-\frac{1}{2},\infty$.
The theory is ill-defined at $C_+<-\frac{1}{2}$, because the energy is not bounded from below.
At $T=T_c$, the coefficient in front of $|u_1|^2$ becomes zero, and hence,
$|u_1|$ can take any value between 0 and $\frac{1}{2}$.

\begin{figure}[htbp]
	\begin{center}
		\rotatebox{0}{
			\includegraphics[width=8cm]{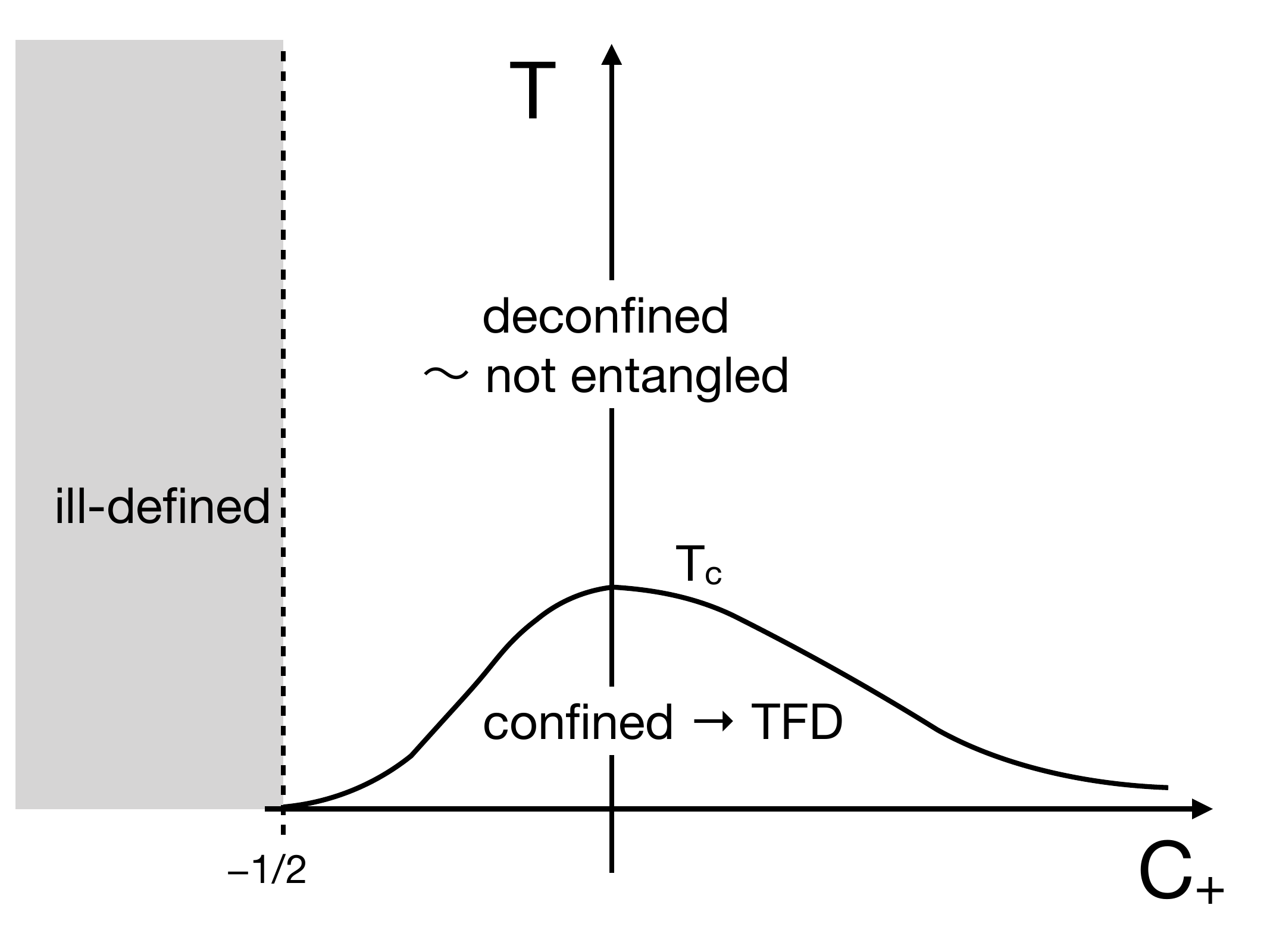}}
	\end{center}
	\caption{The sketch of the phase diagram of the gauged Gaussian matrix model,
		with $(1+2C_+)(1-2C_-)=1$.
	}\label{Fig:Gauged-Gaussian-phase-diagram}
\end{figure}

In the confining phase ($T<T_c$), the system is indistinguishable from the ground state up to the $1/N$ corrections.
Therefore, the confining phase should be the TFD, up to the $1/N$ corrections.
Associated with the deconfinement, the harmonic oscillators are excited, and hence,
the quantum entanglement decreases.
Hence we expect the phase diagram shown in Fig.~\ref{Fig:Gauged-Gaussian-phase-diagram}.
There is one subtlety though; when $T_c$ is small (i.e. $C_+\to-\frac{1}{2}$ or $\infty$),
the jump of the energy is also small, while the entanglement in the ground state is large.
Therefore, the entanglement cannot be washed away immediately. The same holds also when $D$ is large.
It may be an artifact of the free nature of the theory.

\begin{figure}[htbp]
	\begin{center}
		\scalebox{1.0}{
			\rotatebox{0}{
				\includegraphics[width=8cm]{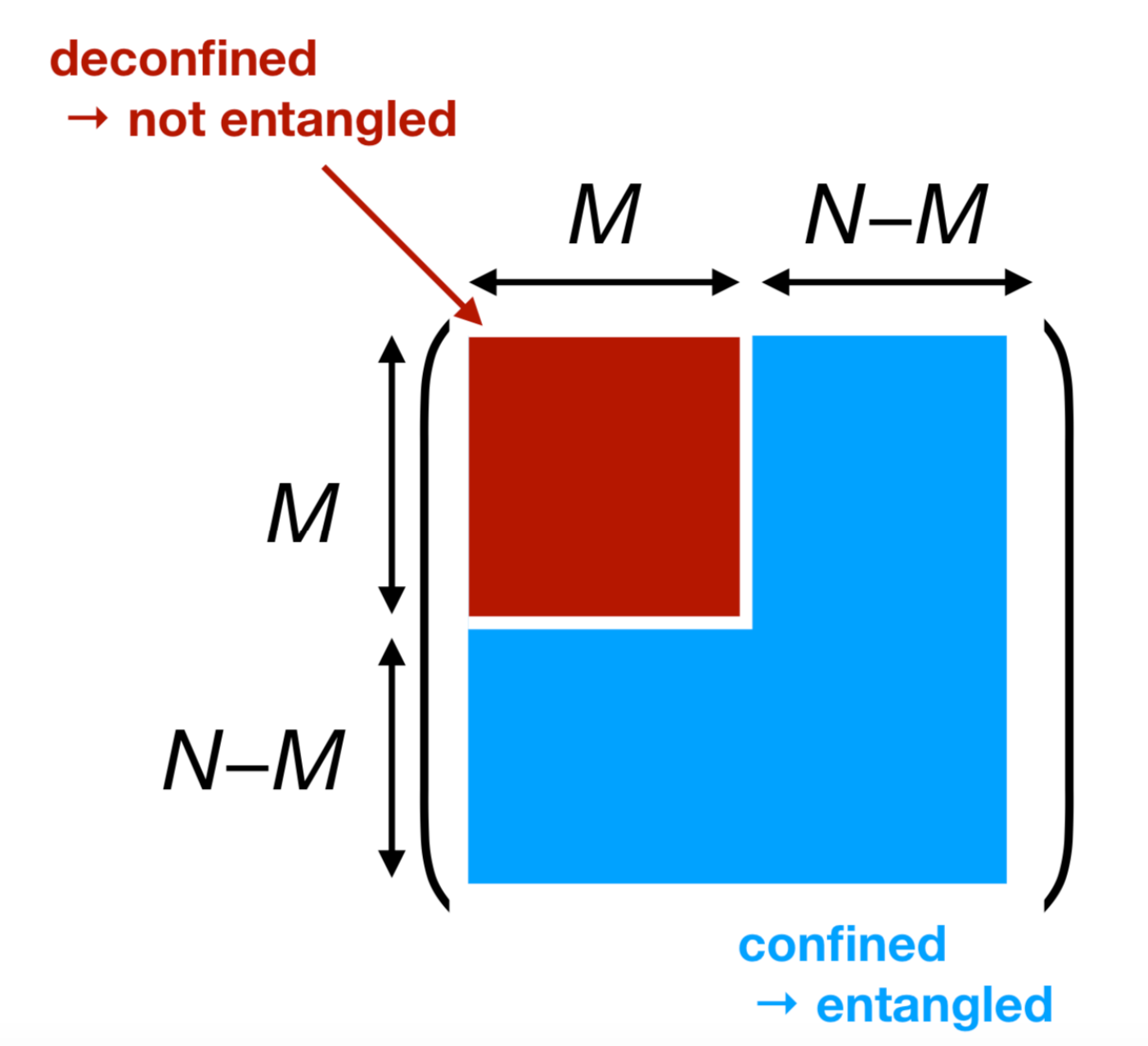}}}
	\end{center}
	\caption{Partial deconfinement phase in the matrix model,
		at small but finite deformation parameter $C_+\ll 1$.
		When the energy is not large enough such that all degrees of freedom are deconfined,
		only part of the degrees of freedom deconfine. The deconfined sectors in the left and right Hilbert spaces are not entangled, while the confined sectors are entangled.
	}\label{Fig:partial_deconfinement_matrix_model}
\end{figure}

Just like the uncoupled theory, we obtain an interesting phase diagram with SU($M$)-deconfinement when we consider the micorocanonical ensemble.  
When the SU($M$) subgroup is deconfined, $N^2-M^2$ degrees of freedom remain confined, namely they remain as the ground state.
Therefore even when the coupling parameter is small, the quantum entanglement survives until all the degrees of freedom deconfine, 
see Fig.\ref{Fig:partial_deconfinement_matrix_model}.
The amount of the entanglement can easily be estimated by counting the number of confined degrees of freedom:
\begin{eqnarray}
\frac{N^2-M^2}{N^2}\times ({\rm entanglement\ entropy\ of\ the\ ground\ state}).
\end{eqnarray}
Here we have assumed that the coupling is small and the entanglement in the deconfined sector is washed away by the thermal excitation.
Because $E\propto M^2$, we can also express the amount of the entanglement as
\begin{eqnarray}
\left(
1-\frac{E}{E_{\rm deconf}}
\right)
\times ({\rm entanglement\ entropy\ of\ the\ ground\ state}),
\end{eqnarray}
where $E_{\rm deconf}$ is the energy needed for the complete deconfinement.

Note that the estimate above is different from the naive entanglement entropy at finite energy;  
we have omitted the contamination coming from the nonzero entanglement entropy in the deconfined sector
which does not correctly measure the quantum entanglement.
In Sec.~\ref{sec:geometric-interpretation}, we will consider the dual gravity description based on this estimate.

\subsubsection*{Gauge-invariance in terms of Hilbert space}
The states in Hilbert space have to satisfy the gauge-singlet constraint. 
This is automatic if we construct the states by acting gauge-invariant operators to the gauge-invariant vacuum. 

The gauge transformation is generated by 
\begin{eqnarray}
\hat{G}_\alpha=i\sum_{I,\beta,\gamma}f_{\alpha\beta\gamma}
\left(
\hat{X}_I^\beta\hat{P}_{X,I}^\gamma
+
\hat{Y}_I^\beta\hat{P}_{Y,I}^\gamma
\right),
\end{eqnarray}
where $f_{\alpha\beta\gamma}$ is the structure constant of the SU($N$) algebra. 
By using 
\begin{eqnarray}
\hat{A}_{{\rm L}I}^\dagger=\frac{\omega\hat{X}_I-i\hat{P}_{X,I}}{\sqrt{2\omega}}, 
\qquad
\hat{A}_{{\rm L}I}=\frac{\omega\hat{X}_I+i\hat{P}_{X,I}}{\sqrt{2\omega}}, 
\nonumber\\
\hat{A}_{{\rm R}I}^\dagger=\frac{\omega\hat{Y}_I-i\hat{P}_{Y,I}}{\sqrt{2\omega}}, 
\qquad
\hat{A}_{{\rm R}I}=\frac{\omega\hat{Y}_I+i\hat{P}_{Y,I}}{\sqrt{2\omega}}, 
\end{eqnarray}
we can rewrite it as
\begin{eqnarray}
\hat{G}_\alpha=\sum_{I,\beta,\gamma}f_{\alpha\beta\gamma}
\left(
\hat{A}_{{\rm L}I}^{\dagger\beta}\hat{A}_{{\rm L}I}^\gamma
+
\hat{A}_{{\rm R}I}^{\dagger\beta}\hat{A}_{{\rm R}I}^\gamma
\right). 
\end{eqnarray}
Hence the ground state of the uncoupled theory $|0\rangle_{\rm L}|0\rangle_{\rm R}$, 
or more explicitly $\otimes_{I,\alpha}\left(|0\rangle_{{\rm L}I\alpha}|0\rangle_{{\rm R}I\alpha}\right)$, 
is gauge-invariant: 
\begin{eqnarray}
\hat{G}_\alpha\left(|0\rangle_{\rm L}|0\rangle_{\rm R}\right)=0.
\end{eqnarray} 

By using 
\begin{eqnarray}
\hat{A}^\dagger_{\pm I}
=
\frac{r_\pm+r_\pm^{-1}}{2\sqrt{2}}\left(
\hat{A}_{{\rm L}I}^\dagger
\pm
\hat{A}_{{\rm R}I}^\dagger
\right)
-
\frac{r_\pm-r_\pm^{-1}}{2\sqrt{2}}\left(
\hat{A}_{{\rm L}I}
\pm
\hat{A}_{{\rm R}I}
\right),
\end{eqnarray}
we obtain 
\begin{eqnarray}
\hat{G}_\alpha
\propto
\sum_{I,\beta,\gamma}f_{\alpha\beta\gamma}
\left(
\hat{A}_{+I}^{\dagger\beta}\hat{A}_{+I}^\gamma
+
\hat{A}_{-I}^{\dagger\beta}\hat{A}_{-I}^\gamma
\right). 
\end{eqnarray}
Hence the ground state of the coupled Hamiltonian $|0\rangle_{+}|0\rangle_{-}$ is also gauge-invariant. 
\subsubsection{Yang-Mills Matrix Model}\label{sec:BMN-matrix-model}
As a concrete example with interaction, let us consider the Yang-Mills matrix model. 
The action of the uncoupled model we consider is
\begin{eqnarray}
S
&=&
N\int_0^{\beta} dt {\rm Tr}
\left(
\frac{1}{2}(D_tX_I)^2
-
\frac{1}{4}[X_I,X_J]^2
\right),
\end{eqnarray}
where $I$ and $J$ run from 1 to $D$. 
The simulation data \cite{Bergner:2019rca} is consistent with the partial deconfinement.
The coupled version is 
\begin{eqnarray}
S
&=&
N\int_0^{\beta} dt {\rm Tr}
\left(
\frac{1}{2}(D_tX_I)^2
-
\frac{1}{4}[X_I,X_J]^2
\right)
+
N\int_0^{\beta} dt {\rm Tr}
\left(
\frac{1}{2}(D_tY_I)^2
-
\frac{1}{4}[Y_I,Y_J]^2
\right)
\nonumber\\
& &
\qquad
+
\frac{NC_{I+}}{2}\int_0^{\beta} dt {\rm Tr}
\left(
X_I+Y_I
\right)^2
-
\frac{NC_{I-}}{2}\int_0^{\beta} dt {\rm Tr}
\left(
X_I-Y_I
\right)^2.
\label{action:gauged_coupled_BFSS}
\end{eqnarray}

We can take only $C_{I+}\ge 0$ and $C_{I-}\le 0$,
because otherwise the potential is not bounded from below
(if we introduce the mass term, for example by considering the plane-wave deformation \cite{Berenstein:2002jq},
$C_{I+}< 0$ and $C_{I-}> 0$ are allowed).
We expect that the X-Y coupling introduces entanglement between X and Y sectors, 
although it is not easy to see analytically if the TFD naturally appears or not.
Below the deconfinement transition there is no temperature dependence up to the $1/N$-corrections,
and hence, the entanglement should survive up to the deconfinement temperature.
When the X-Y coupling is not too large, the jump of the energy at deconfinement transition should be large enough to eliminate the entanglement in the ground state.
Note that, as in the case of the Gaussian matrix model, the entanglement can survive until all the degrees of freedom deconfine.

\subsection{Coupled vector model}\label{sec:vector_model}
In addition to the Gaussian matrix model discussed in Sec.~\ref{sec:Gauged-Gaussian-MM}, 
another analytically-solvable example is the coupled free vector model on S$^1\times$S$^2$, with the gauge-singlet constraint.
We introduce $N$-component vectors $\vec{\phi}_f$, where $f=1,\cdots,N_f$ are the flavor index.
Hence the Lagrangian density is ${\cal L}=\frac{1}{2}\sum_{f=1}^{N_f}(\partial_\mu\vec{\phi}_f)^2$. 
In the path integral formalism, the singlet constraint can be imposed by introducing the Chern-Simons gauge field \cite{Shenker:2011zf}. 
Non-constant modes along the sphere have bigger `effective mass' $\sim\sqrt{m^2+\Delta^2}$,
where $m$ is the mass and $\Delta^2$ is the eigenvalue of the Laplacian.
Due to the lack of the interaction, each mode behaves as harmonic oscillator, whose frequency is the same as the `effective mass'.
Therefore we can use the findings of Sec.~\ref{sec:coupled-harmonic-oscillators}. 
The left-right coupling $C\sum_{f=1}^{N_f}\left(\vec{\phi}_{{\rm L},f}\cdot\vec{\phi}_{{\rm R},f}\right)$ makes each pair the TFD with certain shifted frequency and effective temperature.
Note that the values of the shifted frequency and effective temperature depend on the pair.
The ground state is the tensor product of such TFDs.

When $m$ and coupling parameters are much smaller than the curvature of the sphere,
the phase structure is close to the massless theory with $2N_f$ flavors. 
There is a Gross-Witten-Wadia (GWW) type phase transition \cite{Shenker:2011zf} at 
$T=\sqrt{\frac{3N}{2\pi^2 N_f}}$ \cite{Shenker:2011zf}, 
which separates the partially deconfined phase and completely deconfined phase \cite{Hanada:2019czd}. 

\begin{figure}[htbp]
	\begin{center}
		\scalebox{1.0}{
			\rotatebox{0}{
				\includegraphics[width=8cm]{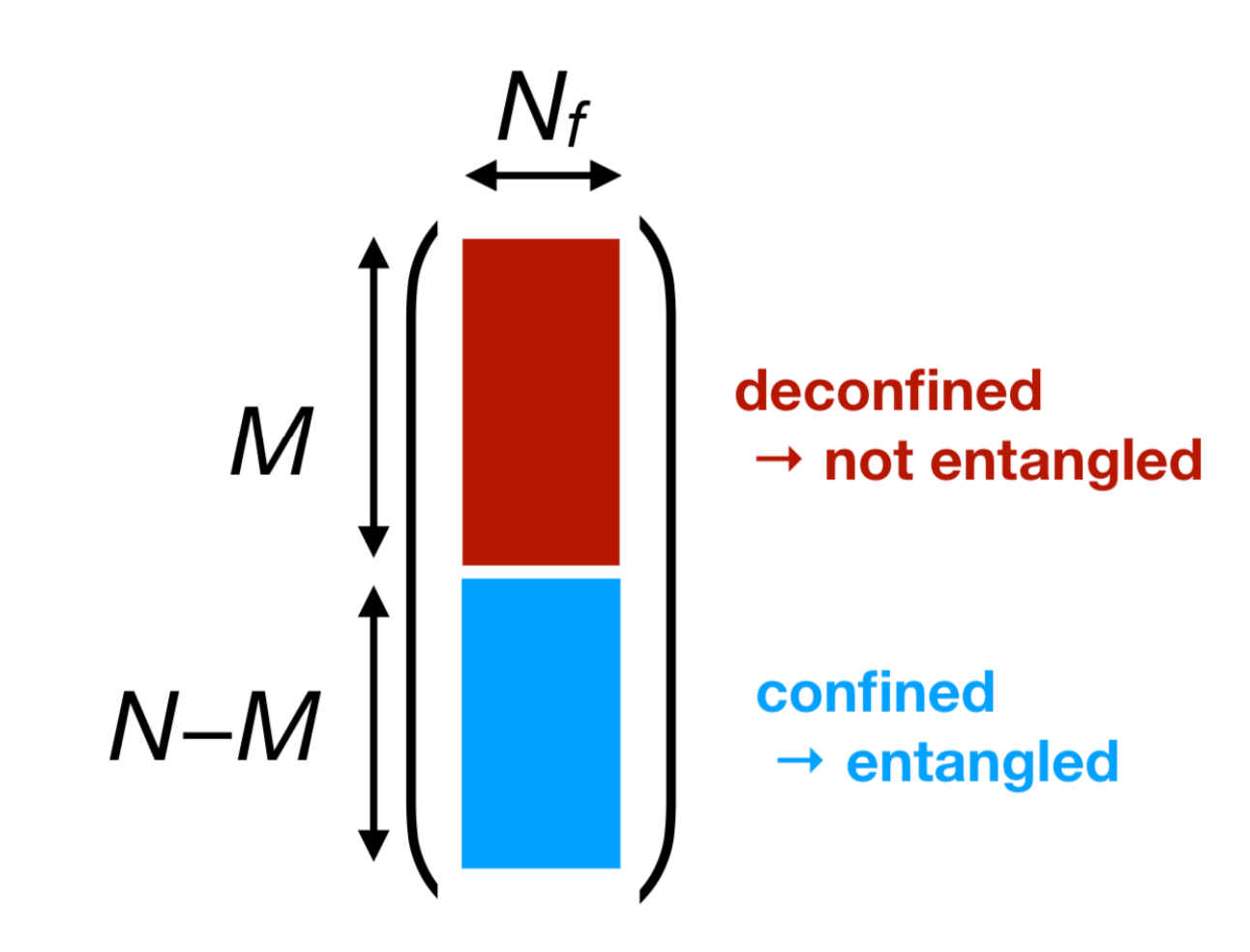}}}
	\end{center}
	\caption{Partial deconfinement phase in the vector model, at small but finite deformation parameter $C_+\ll 1$.
		The deconfined sectors in the left and right Hilbert spaces are not entangled, while the confined sectors are entangled.
	}\label{Fig:partial_deconfinement_vector_model}
\end{figure}

As in the case of the matrix models discussed in Sec.~\ref{sec:matrix-model},  
the entanglement survives all the way up to the GWW transition.
It is rather surprising that the entanglement survives up to such high temperature, which is of order $\sqrt{N}$.
When the temperature is $T=\sqrt{\frac{3M}{2\pi^2 N_f}}$, SU($M$) in SU($N$) is deconfined,
and hence $\frac{M}{N}$ of the entanglement at zero temperature is lost,
see Fig.~\ref{Fig:partial_deconfinement_vector_model}.
The amount of the remaining entanglement is
\begin{eqnarray}
\frac{N-M}{N}\times ({\rm entanglement\ entropy\ of\ the\ ground\ state})\,,
\end{eqnarray}
or equivalently,
\begin{eqnarray}
\left(
1-\frac{T^2}{T_{\rm GWW}^2}
\right)
\times ({\rm entanglement\ entropy\ of\ the\ ground\ state})\ .
\end{eqnarray}

\subsection{Geometric interpretation}\label{sec:geometric-interpretation}
In this section, we propose connections between the mechanism of deconfinement,
loss of the entanglement and the dual gravity interpretation (the disappearance of the traversable wormhole).

As mentioned in Sec.~\ref{sec:matrix-model} and Sec.~\ref{sec:vector_model},
it has been proposed \cite{Hanada:2016pwv,Hanada:2018zxn,Hanada:2019czd,Berenstein:2018lrm}
that the deconfinement transition takes place gradually,
namely there is a partially deconfined phase in which a  part of the SU$(N)$ gauge group, SU$(M)$, is deconfined
and the rest is still confined. The size of the partially deconfined sector $M$ increases gradually toward $N$.
Via AdS/CFT duality, the natural candidate of the gravity dual of the partially deconfined phase is the black hole with negative specific heat,
which is smaller than the AdS scale \cite{Hanada:2016pwv,Hanada:2019czd}. 
While the proposal has been made for the usual, uncoupled gauge theories,
it is natural to expect the same mechanism for the coupled theories, at least when the coupling term is small.
Then what would be the dual gravity interpretation?

The coupled SYK model has a first order transition similar to the deconfinement, and an intermediate state, which resembles the partially deconfined phase, exists \cite{Maldacena:2018lmt}.
The dual gravity interpretation proposed in~\cite{Maldacena:2018lmt} is that
the low-temperature phase ($\sim$ confinement phase, strong entanglement) is dual to the traversable wormhole,
the high-temperature phase ($\sim$ completely deconfined phase, (almost) no entanglement)
is dual to two separate black holes without wormhole,
and the intermediate phase ($\sim$ partially deconfined phase) describes gradual formation of black hole and
disappearance of the traversable wormhole.

If this picture is indeed true for the gauged model, it is confirming the following gravity interpretation \cite{Hanada:2016pwv,Hanada:2018zxn,Hanada:2019czd}:
in the partially deconfined phase, 
the deconfined SU$(M)$-sector describes a small black hole,
while the confined sector describes the rest of the geometry in the dual gravity theory. 
When $\frac{M}{N}\ll 1$, the black hole is so small that the wormhole geometry is not much affected. 
As $M$ increases, the geometry is gradually filled by the black hole and the wormhole becomes thinner. 
In the completely deconfined phase,
all degrees of freedom are excited, and the two copies of deconfined phases are not entangled.
In the language of gravity, the dual black hole is so large that it is almost completely filling AdS
and the traversable wormhole is destroyed.

\section{Conclusion}
In the introduction, we have set two goals of this paper: 
(i) introduce appropriate couplings such that the the ground state of the coupled Hamiltonian mimics TFD, 
and (ii) improve the understanding about the relation between the quantum entanglement and spacetime.  
As for the first goal, we have confirmed the validity of the proposal for the SYK model, 
and generalized it to spin and free-fermion systems as well as bosonic systems including matrix and vector models.
As for the second, we pointed out that the partial deconfinement enables us to estimate the entanglement at finite temperature
and  leads to a natural geometric interpretation.   
The wormhole in the gravity side arises from the entanglement between the color degrees of freedom in the QFT side. 
We regard this mechanism as a good demonstration
of the importance of the entanglement between the color degrees of freedom in the emergence of the bulk geometry
from quantum field theory via holography. 

\begin{center}
	{\Large {\bf Acknowledgements}}
\end{center}
The authors would like to thank Ofer Aharony, Tarek Anous, Micha Berkooz, Matthew Headrick, Ping Gao, Antonio Garc\'{i}a-Garc\'{i}a, Juan Maldacena, Tomoki Nosaka, Rob Pisarski, Dario Rosa, Huajia Wang and Nico Wintergerst for useful discussions. C.~P. thank Shanghai Jiaotong University, the Asia Pacific Center for Theoretical Physics, the Kavli Institute for Theoretical Science at the University of Chinese Academy of Science, Tsinghua University, the Songshan Lake Materials Laboratory, and The Weizmann Institute of Science for warm hospitality during the various stages of the project.    
The work of F.~A. benefited from the support of the project THERMOLOC
ANR-16-CE30-0023-02 of the French National Research
Agency (ANR).   
M.~H.  was supported by the STFC Ernest Rutherford Grant ST/R003599/1
and JSPS  KAKENHI  Grants17K1428.
M.~H. thanks Department of Physics, Brown University for hospitality during his stay 
at the early stage of this work. 
A.~J. and C.~P. were supported by the US Department of Energy under contract DE-SC0010010 Task A. 
C.~P. was also supported by the U.S. Department of Energy grant DE-SC0019480 under the HEP-QIS QuantISED program and by funds from the University of California. C.~P was also supported  by funds from the Kavli Institute for Theoretical Science (KITS) and a startup funding from the University of Chinese Academy of Science (UCAS) during the final stage of this paper for publication.
We acknowledge PRACE for awarding access to HLRS’s Hazel Hen computer based in
Stuttgart, Germany under Grant No. 2016153659, as well as
the use of HPC resources from CALMIP (Grants No. 2018-
P0677, 2019-P0677 and 2020-P0677) and GENCI (Grants No. 2018-
A0030500225 and 2019-A0030500225). Our numerical calculations are based on the PETSc~\cite{Petsc1,Petsc2,Petsc3} and SLEPc~\cite{Slepc} libraries.

\appendix

\section{Methods for numerical calculations}
\label{sec:appendix_numerics}

\subsection{Coupled Spins and SYK models}

For the spin chain and SYK models, we use similar exact diagonalization techniques. 
First, we describe the basis choice for both models. For the SYK model, the size of the total Hilbert space is $2^{N/2}$ for $N$ Majorana fermions. From two Majorana fermions a Dirac fermion can be constructed. Let us use 
$\hat{c}_j = \frac{1}{\sqrt{2}}(\hat{\chi}^j_{\rm L} - i \hat{\chi}^j_{\rm R})$ as a particular choice. 
The total number of Dirac fermions can be measured by the operator 
$\hat{Q}=\sum_{i=1}^{N/2} {\hat{c}^\dagger_i \hat{c}_i}$. 
The fermion parity is defined by $\hat{P}=\frac{1-(-1)^{\hat{Q}}}{2}$. 
Note that $\hat{H}_{\rm int}=\mu\left(\frac{N}{4}-\hat{Q}\right)$. 
It is straightforward to see that $[\hat{H},\hat{P}]=0$. 
Note also that $\hat{Q} \mod 4$ commutes with $\hat{H}$; see e.g. Ref.~\cite{Garcia-Garcia:2019poj}. 
This choice of $\hat{c}_j$ for the Dirac basis is useful for numerical calculations, as explained in Ref.~\cite{Lantagne:2019}, as the Hamiltonian is real for $q=4$ and $q=8$ and block diagonal with respect to $\hat{Q} \mod 4$. Both points ease computations by limiting memory requirements.The ground state of the coupled system is always located in the $Q \mod 4=0$ sector for any $\mu \neq 0$. For the coupled spin chains, the coupling preserves the U(1) symmetry of each chain (the magnetization $\sigma^z_\alpha=\sum_{i=1}^{L/2} \sigma_{i,\alpha}^z$ is conserved for $\alpha=L,R$), and furthermore, we find that the coupled ground state is always in the sector where these two magnetizations are of opposed sign. The total Hilbert space size in this sector is then 
$ \frac{L!}{((L/2)!)^2} \propto \frac{2^L}{\sqrt{L}}$.

To reach systems larger than those accessible with full diagonalization (limited to $L \leq 16$ for the spin chain, $N\leq 32,36$ for SYK), we take advantage of the sparse nature of the matrix representation of the Hamiltonians. Indeed, each matrix line contains only on average $\propto 3L/4$ ($\propto (N/2)^4/192$ in the complex basis) non-zero elements for the coupled spin chain (SYK) model. Note the favorable prefactor for the coupled SYK system compared to a single SYK model which has $\propto (N/2)^4/24$ non-zero matrix elements per line. This overall small number of non-zero matrix elements permits to use sparse linear algebra iterative algorithms such as the Lanczos algorithm to obtain the ground state of the coupled system. We can reach systems of size up to $L=32$ spins for the spin chain (even though we only present data up to $L=28$ for computational resources reasons), and up to $N=52$ for the SYK model. To reach these large systems, we need to use large-scale parallel computations, which is possible with sparse iterative methods. For the computation of the KL divergence in the case of the spin system, we furthermore perform a singular value decomposition (SVD) of the coupled ground state once it has been obtained. For this, we find it beneficial to construct the basis of the coupled system using Lin tables~\cite{Lin:1990} for the right and left subsystems respectively.

The thermofield computation can become also very costly for large $L,N$. Indeed, in the method detailed in Ref.~\cite{Garcia-Garcia:2019poj}, one first need to compute all eigenstates $|E\rangle$ of a single system with $N/2$ particles (which is easy for the considered sizes using full diagonalization) and then perform the outer product between all these eigenstates in Eq.~\ref{eq:defTFD}, which is the most demanding part for large systems, before computing the overlap with the coupled ground state. We do not keep in memory all matrix elements of outer-product but only compute the overlap line-by-line (as explained in Ref.~\cite{Garcia-Garcia:2019poj}). The outer-product is performed using optimized shared-memory parallel routines of the BLAS library. The $U(1)$ symmetry of the spin chain eases this computation as only corresponding sectors of magnetization of the single spin chain (with $L/2$ spins) are matched to form the TFD.

For the coupled SYK model, we did not find an efficient way for computing the TFD double using the full diagonalization of a single SYK model using the real representation of Ref.~\cite{Lantagne:2019} (which is needed to reach large $N$ for the ground state of the coupled model). There we use a different method, also taking advantage of sparsity. As emphasized in Ref.~\cite{Maldacena:2018lmt,Lantagne:2019}, the TFD at $\beta=0$ has a simple representation (a single basis state $|I\rangle = | 00000\rangle$ in the coupled basis of Ref.~\cite{Lantagne:2019}). The TFD at finite $\beta$ can be obtained (up to normalization) by applying $\exp (-\beta H (\mu=0)/4 )$ to this  state. While this requires to use the full coupled Hamiltonian (even though systems are not coupled), we can use Krylov expansion techniques~\cite{Nauts:1983} to perform this application, without the need of forming explicitly the matrix $\exp (-\beta H (\mu=0)/4)$. In practice, all the computations are done in the same run: we start by computing the ground state of $H(\mu)$, store it in memory, and modify the diagonal of the matrix to set $\mu=0$. We then apply recursively $\exp(-\delta \beta H (\mu=0)/4)$ to $|{\rm TFD} (\beta=0)\rangle$ in small steps (typically chosen as $\delta \beta = 1/(100\mu)$). Starting from $\beta=0$ (where the overlap with the coupled ground state is minimal), the overlap grows as $\beta$ increases until it reaches a maximum, which is the value we seek. We stop the iterative application of $\exp(-\delta \beta H(\mu=0) /4 )$ as soon as we observe a drop in the overlap. 

We compute the maximal overlap for each disorder realization (of $h_i$ or $J_{j_1 \cdots j_q}$), then average over disorder to obtain the results presented in the main text. We use from 30 to 1000 realizations of disorder (depending on the system size) for each value of $\mu$.

\subsection{Coupled harmonic oscillators}\label{sec:numerical_calculation_harmonic_oscillator}
We write $\hat{\rho}_{\rm thermal}$ explicitly in the L-R basis, with the explicit cutoff, $n_{\rm L},n_{\rm R},n_\pm<\Lambda$.
Then the dimension of the truncated Hilbert space is $\Lambda^2$.

The only nonzero components of $\hat{a}_{\rm L,R}$ and $\hat{a}_{\rm L,R}^\dagger$ are:
\begin{eqnarray}
\langle n_{\rm L},n_{\rm R}|\hat{a}_{\rm L}|n_{\rm L}+1,n_{\rm R}\rangle
=
\sqrt{n_{\rm L}+1},
\end{eqnarray}
\begin{eqnarray}
\langle n_{\rm L},n_{\rm R}|\hat{a}_{\rm R}|n_{\rm L},n_{\rm R}+1\rangle
=
\sqrt{n_{\rm R}+1},
\end{eqnarray}
\begin{eqnarray}
\langle n_{\rm L},n_{\rm R}|\hat{a}_{\rm L}^\dagger|n_{\rm L}-1,n_{\rm R}\rangle
=
\sqrt{n_{\rm L}},
\end{eqnarray}
\begin{eqnarray}
\langle n_{\rm L},n_{\rm R}|\hat{a}_{\rm R}^\dagger|n_{\rm L},n_{\rm R}-1\rangle
=
\sqrt{n_{\rm R}}.
\end{eqnarray}
By using them, we can construct $\hat{a}_\pm$ and $\hat{a}_\pm^\dagger$ explicitly.
By utilizing the sparseness, we can restrict the cost of the multiplication of $\hat{a}_\pm$ and $\hat{a}_\pm^\dagger$ on a generic state to be $O(\Lambda^2)$.

Note that, if we simply define $|n_{\rm L},n_{\rm R}\rangle=\frac{\hat{a}_{\rm L}^{\dagger n_{\rm L}}\hat{a}_{\rm R}^{\dagger n_{\rm R}}}{\sqrt{n_{\rm L}!n_{\rm R}!}}|0\rangle_{\rm coupled}$,
the norm can deviate from 1 due to the cutoff effect.
We multiplied a positive number so that the norm becomes 1.
They are not exactly orthogonal, due to the regularization effects.

The thermal entropy of the total system can easily be calculated analytically, so we did not use the numerical method
in the results shown in this paper.\footnote{We used it for debugging purposes, namely we checked that the entropy
	of the total system is correctly obtained.}

In order to obtain the entanglement entropy, we constructed the reduced density matrix numerically,
calculated the eigenvalues of the reduced density matrix $p_1,p_2,\cdots,p_\Lambda$,
and then determined the entropy as $-\sum_i p_i\log p_i$. In order to avoid numerical singularities,
we have omitted the eigenvalues smaller than $10^{-15}$.
\subsubsection{Comparison with analytic results}
For the harmonic oscillator, there are simple analytic formulas:
\begin{eqnarray}
Z(T,\omega)
=
\sum_{n=0}^\infty
e^{-\beta\omega(n+\frac{1}{2})}
=
\frac{1}{e^{\frac{1}{2}\beta\omega}-e^{-\frac{1}{2}\beta\omega}},
\end{eqnarray}
\begin{eqnarray}
F(T,\omega)
=
-\frac{\log Z}{\beta}
=
T\log\left(
e^{\frac{1}{2}\beta\omega}-e^{-\frac{1}{2}\beta\omega}
\right),
\end{eqnarray}
\begin{eqnarray}
E(T,\omega)
=
-\frac{\partial\log Z}{\partial\beta}
=
\frac{\omega}{2}
\frac{e^{\frac{1}{2}\beta\omega}+e^{-\frac{1}{2}\beta\omega}}{e^{\frac{1}{2}\beta\omega}-e^{-\frac{1}{2}\beta\omega}},
\end{eqnarray}
\begin{eqnarray}
S(T,\omega)
=
\frac{E-F}{T}.
\end{eqnarray}

Thermal entropy of the coupled harmonic oscillators is
\begin{eqnarray}
S_{\rm thermal}
=
S(T,\omega_+)
+
S(T,\omega_-),
\end{eqnarray}
where
$\omega_+=\sqrt{\omega^2+2C_+}$, $\omega_-=\sqrt{\omega^2-2C_-}$.

To make the ground state to the thermofield double state,
we take $r_+=\frac{1}{r_-}$, where $r_\pm=\sqrt{\frac{\omega_\pm}{\omega}}$.

The entanglement entropy of the ground state is
\begin{eqnarray}
S_{\rm EE,L}
=
S_{\rm EE,R}
=
S(T_{\rm eff},\omega),
\end{eqnarray}
where $T_{\rm eff}$ is defined by
\begin{eqnarray}
T_{\rm eff}
=
-
\frac{\omega}{2\log\left(\left|\frac{r_+-r_+^{-1}}{r_++r_+^{-1}}\right|\right)}.
\end{eqnarray}
$T_{\rm eff}$ is zero when $C_+=0$, and diverges as $C_+\to\infty$ or $C_+\to-\frac{1}{2}$.

\begin{figure}[htbp]
	\begin{center}
		\rotatebox{-90}{
			\includegraphics[width=5cm]{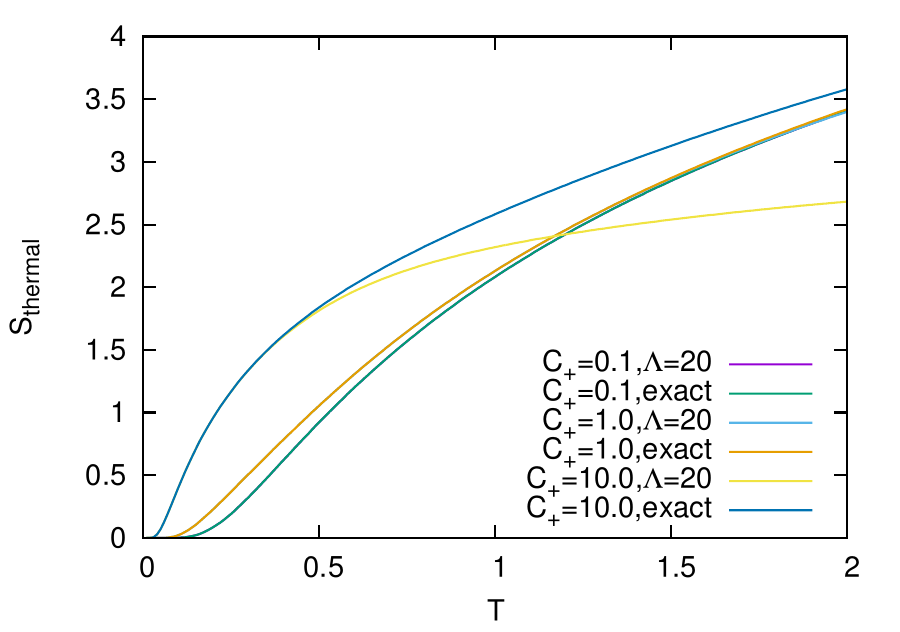}}
	\end{center}
	\caption{Thermal entropy, analytic result vs numerical result, $\omega=1$, with cutoff $\Lambda=20$.
	}\label{fig:sanity_check_coupled_harmonic_oscllators}
\end{figure}

We can use these relations to check the validity of the numerical calculations.
In Fig.~\ref{fig:sanity_check_coupled_harmonic_oscllators}, we have plotted the thermal entropy
calculated numerically and the one obtained analytically. We can observe a good agreement, when the cutoff $\Lambda$ is sufficiently large.

\end{document}